\newcommand{\resection}[1]{\setcounter{equation}{0}\section{#1}}

\documentclass[a4paper,11pt]{article}
\usepackage{amssymb}

\usepackage{amsmath}
\usepackage{epsfig}

\newcommand{\EQ}{\begin{equation}}
\newcommand{\EN}{\end{equation}}
\newcommand{\bea}{\begin{eqnarray}}
\newcommand{\eea}{\end{eqnarray}}

\begin{document}

\setcounter{page}{0} \topmargin0pt \oddsidemargin5mm \renewcommand{%
\thefootnote}{\arabic{footnote}} \newpage \setcounter{page}{0} 
\begin{titlepage}
\begin{flushright}
%OUTP-96-19S\\
SISSA 07/2006/EP
\end{flushright}
\vspace{0.5cm}
\begin{center}
{\large {\bf The composite operator $T\bar{T}$ in sinh-Gordon} \\
{\bf and a series of massive minimal models}}\\ 
\vspace{1.8cm}
{\large Gesualdo Delfino$^a$ and Giuliano Niccoli$^b$} \\
\vspace{0.5cm}
{\em ${}^a$ International School for Advanced Studies (SISSA)}\\
{\em via Beirut 2-4, 34014 Trieste, Italy}\\
{\em INFN sezione di Trieste}\\
{\em e-mail: delfino@sissa.it}\\
{\em ${}^b$ Laboratoire de physique, \'Ecole Normale Sup\'erieure de Lyon}\\
{\em 69364 Lyon, France}\\
{\em e-mail: Giuliano.Niccoli@ens-lyon.fr}
\end{center}
\vspace{1.2cm}

\renewcommand{\thefootnote}{\arabic{footnote}}
\setcounter{footnote}{0}

\begin{abstract}
\noindent
The composite operator $T\bar{T}$, obtained from the components of the energy-momentum tensor, enjoys 
a quite general characterization in two-dimensional quantum field theory also away from criticality. We use
the form factor bootstrap supplemented by asymptotic conditions to determine its matrix elements in the 
sinh-Gordon model. The results extend to the breather sector of the sine-Gordon model and to the minimal 
models $\mathcal{M}_{2/(2N+3)}$ perturbed by the operator $\phi_{1,3}$.
\end{abstract}
\end{titlepage}

\newpage \resection{Introduction} The ability of a quantum field theory to
describe a system with infinitely many degrees of freedom is reflected by an
infinite-dimensional operator space. In two dimensions, the detailed
structure of the operator space at a generic fixed point of the
renormalization group was revealed by the solution of conformal field
theories \cite{BPZ}. It is divided into different operator families, each
one consisting of a primary and infinitely many descendants. Within an
operator family, the scaling dimensions differ from that of the primary by
integer numbers that label different `levels'.

Perturbative arguments lead to the conclusion that this same structure is
maintained when conformal invariance is broken by a perturbation producing a
mass scale \cite{Taniguchi}. If the massive theory is integrable, the
operator space can be studied non-perturbatively within the form factor
bootstrap approach \cite{KW,Smirnov}. It was shown for several models that
the global counting of solutions of the form factor equations matches that
expected from conformal field theory \cite{counting}.

As for the correspondence between solutions of the form factor equations and
operators, asymptotic conditions at high energies play a crucial role. While
primary operators are naturally associated to the solutions with the mildest
asymptotic behavior, we argued in \cite{ttbar} that specific asymptotic
conditions selecting the solutions according to the level can be identified.
These were used in \cite{seven} to show the isomorphism between the critical
and off-critical operator spaces in the Lee-Yang model, level by level up to
level $7$.

Asymptotic conditions, however, cannot determine completely a descendant
operator in a massive theory. Indeed, they leave unconstrained terms which
are subleading at high energies and depend on the way the operator is
defined away from criticality. The operator $T\bar{T}$, obtained from the
components of the energy-momentum tensor, appears as the natural starting
point in relation to the problem of the off-critical continuation of
descendant operators. Indeed, being the lowest non-trivial scalar descendant
of the identity, this operator allows for a quite general characterization
in two-dimensional quantum field theory. A.~Zamolodchikov showed how to
define it away from criticality subtracting the divergences which arise in
the operator product expansion of $T$ and $\bar{T}$ \cite{Sasha}. We showed
in \cite{ttbar} for the massive Lee-Yang model that, with this information,
the form factor programme outlined above allows to uniquely determine $T\bar{%
T}$ up to an additive derivative ambiguity which is intrinsic to this
operator. Our results have been successfully compared with conformal
perturbation theory in \cite{BM}.

In this paper we address the problem of determining $T\bar{T}$ in the
sinh-Gordon model. The essential difference with respect to the Lee-Yang
case is that, while the latter is a minimal model with the smallest operator
content (two operator families), the sinh-Gordon model possesses a
continuous spectrum of primary operators, a circumstance that seriously
complicates the identification of specific solutions of the form factor
equations. The massive Lee-Yang model is the first in the infinite series of
the $\phi_{1,3}$-perturbed minimal models $\mathcal{M}_{2/(2N+3)}$, each one
containing $N+1$ operator families. Due to a well known reduction mechanism 
\cite{LeClair,Smirnovreduction,RS,BLc}, these massive minimal models have to
be recovered from sinh-Gordon under analytic continuation to specific
imaginary values of the coupling. The form factor solution for the operator $%
T\bar{T}$ of the sinh-Gordon model that we construct satisfies this
requirement.

The paper is organized as follows. In the next section we recall a number of
facts about bosonic theories in two dimensions. The form factor solutions
for the primary operators in the sinh-Gordon model are reviewed in
section~3, while the solution for $T\bar{T}$ is constructed in section~4.
Few final remarks are contained in section~5. Six appendices conclude the
paper.

\resection{Bosonic field with a charge at infinity} If $B_{\lambda \mu \nu }$
is a tensor antisymmetric in the first two indices, the energy-momentum
tensor $T_{\mu \nu }$ of a quantum field theory can be modified into 
\begin{equation}
\tilde{T}_{\mu \nu }=T_{\mu \nu }+\partial ^{\lambda }B_{\lambda \mu \nu }
\label{modified}
\end{equation}
preserving the conservation $\partial ^{\mu }\tilde{T}_{\mu \nu }=0$ and the
total energy-momentum $P_{\nu }=\int d\sigma ^{\mu }T_{\mu \nu }$. For a
neutral two-dimensional boson with action 
\begin{equation}
\mathcal{A}=\int d^{2}x\,\left[ \frac{1}{2}\,(\partial _{\mu }\varphi )^{2}+%
\mathcal{V}(\varphi )\right]  \label{A}
\end{equation}
the choice\footnote{%
We denote by $\eta _{\mu \nu }$ the flat metric tensor and by $\epsilon
_{\lambda \mu }$ the unit antisymmetric tensor in two dimensions.} 
\begin{equation}
B_{\lambda \mu \nu }=-\frac{iQ}{\sqrt{2\pi }}\epsilon _{\lambda \mu
}\epsilon _{\nu \rho }\partial ^{\rho }\varphi
\end{equation}
leads to 
\begin{equation}
\partial ^{\lambda }B_{\lambda \mu \nu }=-\frac{iQ}{\sqrt{2\pi }}(\partial
_{\mu }\partial _{\nu }-\eta _{\mu \nu }\Box )\varphi
\end{equation}
and to the variation 
\begin{equation}
\tilde{\Theta}=\Theta +\frac{iQ}{2}\,\sqrt{\frac{\pi }{2}}\,\Box \varphi
\label{thetatilde}
\end{equation}
in the trace of the energy-momentum tensor 
\begin{equation}
\Theta =\frac{\pi }{2}\,T_{\mu }^{\mu }\,\,.  \label{trace}
\end{equation}
The canonical definition $T_{\mu }^{\mu }=-2\mathcal{V}$ and the equation of
motion $\Box \varphi =\partial \mathcal{V}/\partial \varphi $ give the
classical result 
\begin{equation}
\tilde{\Theta}_{cl}=\pi \left( -1+\frac{iQ}{\sqrt{8\pi }}\frac{\partial }{%
\partial \varphi }\right) \mathcal{V}(\varphi )\,\,.  \label{thetacl}
\end{equation}

The parameter $Q$ is dimensionless and goes under the name of ``background
charge'' or ``charge at infinity''. It does not change the particle dynamics%
\footnote{%
In particular, $Q$ does not enter the perturbative calculations based on the
Lagrangian.} but, inducing a modification of the energy-momentum tensor,
essentially affects the scaling properties of the theory.

\vspace{0.3cm} \noindent \textbf{Free massless case.} In the free massless
case corresponding to the action %${\cal V(\varphi)}=0$
\begin{equation}
\mathcal{A}_{0}=\frac{1}{2}\int d^{2}x\,(\partial _{\mu }\varphi )^{2}
\label{A0}
\end{equation}
a nonvanishing $Q$ leaves the energy-momentum tensor traceless ($\tilde{%
\Theta}_{cl}=\tilde{\Theta}=\mathcal{V}=0$) and the theory conformally
invariant. The central charge is 
\begin{equation}
C=1-6Q^{2}  \label{C}
\end{equation}
and the scaling dimension of the primary operators 
\begin{equation}
V_{\alpha }(x)=e^{i\sqrt{8\pi }\,\alpha \,\varphi (x)}  \label{vertex}
\end{equation}
is 
\begin{equation}
X_{\alpha }=2\alpha (\alpha -Q)\,.  \label{X}
\end{equation}
A derivation of these results within the formalism of this paper is given in
appendix~A.

\vspace{.3cm} \noindent \textbf{Minimal models.} For real values of the
background charge the bosonic model can be used to reproduce the minimal
models of conformal field theory with central charge smaller than 1 \cite{DF}%
. Indeed, the requirement that the $2k$-point conformal correlator of an
operator $V_\alpha\sim V_{Q-\alpha}$ is nonvanishing for any positive $k$
selects the values 
\begin{equation}
\alpha=\alpha_{m,n}=\frac12\,[(1-m)\alpha_+ +(1-n)\alpha_-]\,, \hspace{1cm}%
m,n=1,2,\ldots  \label{alphamn}
\end{equation}
with 
\begin{equation}
\alpha_\pm=\frac{Q\pm\sqrt{Q^2+4}}{2}\,\,.
\end{equation}
Equations (\ref{C}) and (\ref{X}) then reproduce the central charge 
\begin{equation}
C_{p/p^{\prime}}=1-6\,\frac{(p-p^{\prime})^2}{pp^{\prime}}  \label{Cminimal}
\end{equation}
and the scaling dimensions 
\begin{equation}
X_{m,n}=\frac{(p^{\prime}m-pn)^2-(p-p^{\prime})^2}{2pp^{\prime}}
\label{Xminimal}
\end{equation}
of the primary operators $\phi_{m,n}$ in the minimal models $\mathcal{M}%
_{p/p^{\prime}}$ \cite{BPZ} through the identification 
\begin{equation}
\alpha_\pm=\pm\left(\frac{p^{\prime}}{p}\right)^{\pm\frac12}\,\,.
\end{equation}
Since $\alpha_+\alpha_-=-1$, one obtains the correspondence 
\begin{equation}
\phi_{m,n}\sim V_{\alpha_{m,n}}=\exp\left\{\frac{i}{2}\left[(m-1)\frac{1}{
\alpha_-}-(n-1)\alpha_-\right]\varphi\right\}\,\,.  \label{phimn}
\end{equation}

It is well known that, although genuine minimal models (i.e. those
possessing a finite number of conformal families which form an operator
space closed under operator product expansion) correspond to rational values
of $p/p^{\prime }$, the above formul\ae ~in fact apply to the degenerate
operators of conformal field theory for continuous values of central charge
smaller than 1.

\vspace{.3cm} \noindent \textbf{Liouville theory.} A deformation of (\ref{A0}%
) which does not introduce any dimensional parameter is obtained adding an
operator which is marginal in the renormalization group sense, namely has $%
X_\alpha=2$. This requirement selects $V_{\alpha_-}\sim
V_{Q-\alpha_-}=V_{\alpha_+}$. To be definite we take 
\begin{equation}
\mathcal{A}_L= \int d^2x\,\left[\frac12\,(\partial_\nu\varphi)^2+ \mu\,e^{%
\sqrt{8\pi}\,b\,\varphi}\right]\,,  \label{AL}
\end{equation}
where we defined 
\begin{equation}
b=i\alpha_-  \label{b}
\end{equation}
and $\mu$ is a coupling constant. For real values of $b$ this is the action
of Liouville field theory, which is conformal and has been extensively
studied in the literature (see e.g. \cite{ZZliouville} for a list of
references). Notice that the condition $\tilde{\Theta}_{cl}=0$ gives the
value $Q_{cl}=-i/b$, which coincides with the exact result 
\begin{equation}
Q=\alpha_++\alpha_-=-i\left(b+\frac1b\right)  \label{Q}
\end{equation}
in the classical limit $b\to 0$. The central charge and scaling dimensions
of exponential operators in Liouville field theory are given by (\ref{C})
and (\ref{X}) with $Q$ given by (\ref{Q}).

\vspace{.3cm} \noindent \textbf{Sinh-Gordon model.} The sinh-Gordon model is
defined by the action 
\begin{equation}
\mathcal{A}_{shG}=\int d^2x\,\left[\frac12\,(\partial_\nu\varphi)^2+ \mu\,e^{%
\sqrt{8\pi}\,b\,\varphi}+\mu^{\prime}\,e^{-\sqrt{8\pi}\,b\,\varphi}\right]\,,
\label{Ashg}
\end{equation}
which can be regarded as a perturbed conformal field theory in two different
ways.

The first one consists in seeing it as a deformation of the Gaussian fixed
point, i.e. the conformal theory with $C=1$. This amounts to setting $Q=0$
keeping $b$ as a free parameter. In such a case, both the exponentials
appearing in the action have scaling dimension $-2b^{2}$ and are never
marginal for real values of $b$. They play a symmetric role and the theory
is invariant\footnote{%
In this case the couplings $\mu $ and $\mu ^{\prime }$ have the same
dimension and can be made equal shifting the field.} under the
transformation $\varphi \rightarrow -\varphi $. The trace of the
energy-momentum tensor, being proportional to the operator which breaks
conformal invariance, is $\tilde{\Theta}=\Theta \sim \mu \cosh \sqrt{8\pi }%
b\varphi $.

The second point of view consists in looking at (\ref{Ashg}) as the
perturbation of the Liouville conformal theory (\ref{AL}) by the operator $%
e^{-\sqrt{8\pi}\,b\,\varphi}$ with scaling dimension $X_{ib}=-2(2b^2+1)$.
Since this is now the operator which breaks conformal invariance, we have $%
\tilde{\Theta}\sim\mu^{\prime}\,e^{-\sqrt{8\pi}\,b\,\varphi}$, a result
which agrees with the classical expectation (\ref{thetacl}) once one uses $%
Q_{cl}$ for $Q$.

%In both cases the action (\ref{Ashg}) contains a dimensionful coupling 
%conjugated to a relevant perturbing operator, so that the theory is massive.

\vspace{.3cm} \noindent \textbf{Sine-Gordon model and its reductions.} The
sine-Gordon action 
\begin{equation}
\mathcal{A}_{sG}=\int d^2x\,\left[\frac12\,(\partial_\nu\varphi)^2-
2\mu\,\cos{\sqrt{8\pi}\,\beta\,\varphi}\right]  \label{Asg}
\end{equation}
can be obtained from (\ref{Ashg}) taking $\mu=\mu^{\prime}$ and 
\begin{equation}
\beta=-ib\,\,.  \label{beta}
\end{equation}

For real values of $\beta$ the only direct interpretation of this action as
a perturbed conformal field theory is as a deformation of the $C=1$
conformal theory through the operators $V_\beta$ and $V_{-\beta}$ with
scaling dimension $2\beta^2$. The perturbation is relevant and the theory is
massive for $\beta^2<1$. In this range the sine-Gordon model is known to be
integrable and its factorized $S$-matrix is known exactly \cite{ZZ}. The
particle spectrum consists of the soliton $A$ and antisoliton $\bar{A}$ and,
in the attractive range $0<\beta^2<1/2$, of their neutral bound states, the
breathers $B_n$ with 
\begin{equation}
1\leq n<\mbox{Int}\left(\frac{\pi}{\xi}\right)  \label{nb}
\end{equation}
and masses 
\begin{equation}
m_n=2M\,\sin\frac{n\xi}{2}\,;  \label{m_n}
\end{equation}
here $\mbox{Int}(x)$ denotes the integer part of $x$, $M$ is the mass of the
soliton and 
\begin{equation}
\xi=\frac{\pi\beta^2}{1-\beta^2}\,\,.  \label{xi}
\end{equation}

Taking $Q\neq 0$ and looking at the sine-Gordon model as a perturbation of
the conformal theory (\ref{AL}) is problematic because the action (\ref{AL})
becomes complex when $b$ is imaginary. Formally, however, this point of view
leads, through the identity 
\begin{equation}
\beta=\alpha_-=-\sqrt{\frac{p}{p^{\prime}}}\,,  \label{betaalpha}
\end{equation}
to the conformal field theories with central charge (\ref{Cminimal})
perturbed by the operator $V_{-\beta}\sim\phi_{1,3}$, namely to the action 
\begin{equation}
\mathcal{A}_{\mathcal{M}_{p/p^{\prime}}}+\lambda\int d^2x\,\phi_{1,3}(x)\,\,.
\label{Arsg}
\end{equation}
This $\phi_{1,3}$-perturbation of the $C<1$ conformal field theories is
known to be integrable for any value of $\lambda$ \cite{Taniguchi} and
massive for a suitable choice of the sign of $\lambda$ \cite{cth}. This
choice is implied in (\ref{Arsg}).

The relation between the sine-Gordon model and the action (\ref{Arsg})
suggested by these formal reasonings can be confirmed and put on firmer
grounds within a framework known as quantum group reduction \cite{LeClair,
Smirnovreduction,RS,BLc}. This relies on the fact that the sine-Gordon $S$%
-matrix commutes with the generators of the affine quantum group $SL(2)_q$
with $q=\exp(i\pi/\beta^2)$, and that for rational values of $%
\beta^2=p/p^{\prime}$ a restriction can be operated in the space of particle
states and operators of the model which is consistent with this algebraic
structure and preserves locality. The quantum field theories obtained
through this reduction mechanism indeed coincide with the perturbed minimal
models (\ref{Arsg}).

While soliton and antisoliton transform as a doublet under the action of the
quantum group, the breathers are scalars. This is why the reduction takes
its simplest form when the space of states can be restricted to the breather
sector. Let us recall that the amplitude for the scattering between the
breathers $B_{m}$ and $B_{n}$ in the sine-Gordon model is\footnote{%
The rapidity variables $\theta _{i}$ parameterize energy and momentum of a
particle as $(p_{i}^{0},p_{i}^{1})=(m\cosh \theta _{i},m\sinh \theta _{i})$, 
$m$ being the mass. The scattering amplitudes depend on the rapidity
difference between the colliding particles.} \cite{ZZ} 
\begin{equation}
S_{mn}(\theta )=t_{(m+n)\frac{\xi }{2\pi }}(\theta )t_{|m-n|\frac{\xi }{2\pi 
}}(\theta )\prod_{j=1}^{\min (m,n)-1}t_{(|m-n|+2j)\frac{\xi }{2\pi }%
}^{2}(\theta )\,,  \label{amplitudes}
\end{equation}
where 
\begin{equation}
t_{\alpha }(\theta )=\frac{\tanh \frac{1}{2}(\theta +i\pi \alpha )}{\tanh 
\frac{1}{2}(\theta -i\pi \alpha )}\,\,.
\end{equation}
While the double poles are associated to multiscattering processes \cite{CT,
Goebel}, the simple poles located at $\theta =i(m+n)\xi /2$ and $\theta
=i(\pi -|m-n|\xi /2)$ correspond to the bound states $B_{n+m}$ and $%
B_{|m-n|} $, respectively, propagating in the scattering channel $B_{m}B_{n}$%
. To be more precise, the first class of simple poles can be associated to
the bound states $B_{m+n}$ only for values of $\xi $ such that $m+n<%
\mbox{Int}(\pi /\xi )$. Indeed, (\ref{nb}) shows that outside this range the
particle $B_{m+n}$ is not in the spectrum of the model in spite of the fact
that the pole in the amplitude (\ref{amplitudes}) may still lie in the
physical strip $\mbox{Im}\,\theta \in (0,\pi )$. In this case, however, this
pole can be explained in terms of a multiscattering process involving
solitons as intermediate states. This is why, for generic values of $\xi $,
the breather sector of the sine-Gordon model is not a self-contained
bootstrap system.

The situation becomes different when $\xi$ takes the special values 
\begin{equation}
\xi_N=\frac{2\pi}{2N+1}\,,\hspace{1cm}N=1,2,\ldots,\,\,\,.  \label{xiN}
\end{equation}
In this case (\ref{nb}), (\ref{m_n}) and (\ref{amplitudes}) show that there
are $N$ breathers and that the formal identities 
\begin{eqnarray}
&& m_n = m_{2N+1-n} \\
&& S_{mn}(\theta) = S_{m,2N+1-n}(\theta) \,  \label{periodicity}
\end{eqnarray}
hold, so that the pole discussed above can be associated to $B_{m+n}$ for $%
m+n\leq N$ and to $B_{2N+1-m-n}$ for $m+n>N$, without any need to resort to
the solitons. Hence, for the values (\ref{xiN}) of the coupling, the
breather sector of the sine-Gordon model provides alone a self-consistent
factorized scattering theory and, consequently, defines an infinite series
of massive integrable models labeled by the positive integer $N$. Equations (%
\ref{xi}) and (\ref{betaalpha}) then identify these massive models with the
minimal models $\mathcal{M}_{2/(2N+3)}$ with central charge 
\begin{equation}
C_N=1-3\,\frac{(2N+1)^2}{2N+3}\,,  \label{CN}
\end{equation}
perturbed by the operator $\phi_{1,3}$ with scaling dimension 
\begin{equation}
X_{1,3}^{(N)}=-2\,\frac{2N-1}{2N+3}\,\,.  \label{XN}
\end{equation}
This conclusion was first reached in \cite{Smirnovreduction}. The
thermodynamic Bethe ansatz \cite{TBA,KM} confirms that the scattering theory
(\ref{amplitudes}) gives the central charges (\ref{CN}) for $\xi=\xi_N$, a
result that can be regarded as a non-perturbative confirmation of the fact
that the charge at infinity does not affect the dynamics of the particles.
The minimal models $\mathcal{M}_{2/(2N+3)}$ possess the $N$ non-trivial
primary fields $\phi_{1,k}$, $k=2,\ldots,N+1$, plus the identity $\phi_{1,1}$%
. The negative values of the conformal data (\ref{CN}) and (\ref{XN}) show
that these models do not satisfy reflection positivity. The case $N=1$
corresponds to the Lee-Yang model \cite{YL,Fisher,Cardy}, the simplest
interacting quantum field theory. Its $S$-matrix $\left.S_{11}(\theta)%
\right|_{\xi=2\pi/3}= t_{2/3}(\theta)$ was identified in \cite{CMyanglee}.

\resection{Primary operators in the sinh-Gordon model} Most of the results
discussed for the sine-Gordon model apply to the sinh-Gordon model (\ref
{Ashg}) through the correspondence (\ref{beta}). In particular also the
latter model is a massive integrable quantum field theory. The particle $B$
interpolated by the scalar field corresponds to the sine-Gordon lightest
breather $B_1$. The scattering amplitude 
\begin{equation}
S(\theta)\equiv S_{11}(\theta)=t_{\frac\xi\pi}(\theta)  \label{Sshg}
\end{equation}
does not possess poles in the physical strip when $b$ is real and completely
specifies the $S$-matrix of the sinh-Gordon model. This amplitude was
proposed and checked against perturbation theory in $b$ in \cite{AK,VG,STW}.
It should be clear from the discussion of the previous section that the $S$%
-matrix is the same for the two ultraviolet limits (Gaussian fixed point and
Liouville theory) compatible with the action (\ref{Ashg}).

The $S$-matrix determines the basic equations satisfied by the matrix
elements of a local operator $\Phi (x)$ on the asymptotic multiparticle
states \cite{KW,Smirnov}. The form factors\footnote{%
We denote by $|0\rangle $ the vacuum state.} 
\begin{equation}
F_{n}^{\Phi }(\theta _{1},\ldots ,\theta _{n})=\langle 0|\Phi (0)|B(\theta
_{1})\ldots B(\theta _{n})\rangle  \label{ff}
\end{equation}
obey the equations 
\begin{align}
& F_{n}^{\Phi }(\theta _{1}+\alpha ,\ldots ,\theta _{n}+\alpha )=e^{s_{\Phi
}\alpha }F_{n}^{\Phi }(\theta _{1},\ldots ,\theta _{n})  \label{fn0} \\
& F_{n}^{\Phi }(\theta _{1},\ldots ,\theta _{i},\theta _{i+1},\ldots ,\theta
_{n})=S(\theta _{i}-\theta _{i+1})\,F_{n}^{\Phi }(\theta _{1},\ldots ,\theta
_{i+1},\theta _{i},\ldots ,\theta _{n})  \label{fn1} \\
& F_{n}^{\Phi }(\theta _{1}+2i\pi ,\theta _{2},\ldots ,\theta
_{n})=F_{n}^{\Phi }(\theta _{2},\ldots ,\theta _{n},\theta _{1})  \label{fn2}
\\
& \mbox{Res}_{\theta ^{\prime }=\theta +i\pi }\,F_{n+2}^{\Phi }(\theta
^{\prime },\theta ,\theta _{1},\ldots ,\theta _{n})=i\left[
1-\prod_{j=1}^{n}S(\theta -\theta _{j})\right] F_{n}^{\Phi }(\theta
_{1},\ldots ,\theta _{n})  \label{fn4}
\end{align}
where the euclidean spin $s_{\Phi }$ is the only operator-dependent
information.

The solutions of the equations (\ref{fn0})-(\ref{fn4}) can be parameterized
as \cite{FMS,KlM} 
\begin{equation}
F_{n}^{\Phi }(\theta _{1},\ldots ,\theta _{n})=U_{n}^{\Phi }(\theta
_{1},\ldots ,\theta _{n})\prod_{i<j}\frac{\mathcal{F}(\theta _{i}-\theta
_{j})}{\cosh \frac{\theta _{i}-\theta _{j}}{2}}\,\,,  \label{fn}
\end{equation}
Here the factors in the denominator introduce the annihilation poles
prescribed by (\ref{fn4}), and 
\begin{equation}
\mathcal{F}(\theta)=\mathcal{N}(\xi)\,\exp\left[2\int_0^\infty\frac{dt}{t}\,
q_\xi(t)\,\frac{\sinh\frac{t}{2}}{\sinh^2 t}\, \sin ^{2}\frac{(i\pi -\theta
)t}{2\pi }\right]  \label{Fmin}
\end{equation}
with 
\begin{equation}
q_\xi(t)=-4\sinh\frac{\xi t}{2\pi}\,\sinh\left[\left(1+\frac{\xi}{\pi}%
\right) \frac{t}{2}\right]
\end{equation}
\begin{equation}
\mathcal{N}(\xi)=\mathcal{F}(i\pi)=\exp\left[-\int_0^\infty\frac{dt}{t}\,
q_\xi(t)\,\frac{\sinh\frac{t}{2}}{\sinh^2 t}\right]\,\,.
\end{equation}
The function $\mathcal{F}(\theta)$ is the solution of the equations 
\begin{equation}
\mathcal{F}(\theta)=S(\theta)\mathcal{F}(-\theta)
\end{equation}
\begin{equation}
\mathcal{F}(\theta+2i\pi)=\mathcal{F}(-\theta)
\end{equation}
with asymptotic behavior 
\begin{equation}
\lim_{|\theta |\rightarrow \infty }\mathcal{F}(\theta)=1\,;
\end{equation}
it also satisfies the functional relation 
\begin{equation}
\mathcal{F}(\theta +i\pi)\mathcal{F}(\theta)=\frac{\sinh\theta } {\sinh
\theta -\sinh i\xi } \,\,.
\end{equation}
The expression (\ref{Fmin}) is convergent in the range $-\pi<\xi<0$ which is
relevant for the sinh-Gordon model.

All the information about the operator is contained in the functions $%
U_{n}^{\Phi }$. They must be entire functions of the rapidities, symmetric
and (up to a factor $(-1)^{n-1}$) $2\pi i$-periodic in all $\theta _{j}$. We
write them in the form 
\begin{equation}
U_{n}^{\Phi }(\theta _{1},..,\theta _{n})=\mathcal{H}_{n}\left( \frac{1}{%
\sigma _{n}^{(n)}}\right) ^{\left( n-1\right) /2}Q_{n}^{\Phi }(\theta
_{1},..,\theta _{n})  \label{un}
\end{equation}
using the symmetric polynomials generated by 
\begin{equation}
\prod_{i=1}^{n}(x+x_{i})=\sum_{k=0}^{n}x^{n-k}\sigma _{k}^{(n)}(x_{1},\ldots
,x_{n})  \label{sigma-def}
\end{equation}
with $x_{i}\equiv e^{\theta _{i}}$, and choosing the constants 
\begin{equation}
\mathcal{H}_{n}=\left(\frac{-8\sin\xi}{2^n\mathcal{F}(i\pi)}%
\right)^{n/2}\,\,.
\end{equation}
The equations (\ref{fn0})--(\ref{fn4}) imply 
\begin{eqnarray}
&&Q_{n}^{\Phi }\left( \theta _{1}+\alpha ,..,\theta _{n}+\alpha \right)
=e^{\left( s_{\Phi }+\frac{n\left( n-1\right) }{2}\right) \alpha
}\,Q_{n}^{\Phi }\left( \theta _{1},..,\theta _{n}\right)  \label{q1} \\
&&Q_{n}^{\Phi }\left( \theta _{1},..,\theta _{i},\theta _{i+1},..,\theta
_{n}\right) =Q_{n}^{\Phi }\left( \theta _{1},..,\theta _{i+1},\theta
_{i},..,\theta _{n}\right)  \label{q2} \\
&&Q_{n}^{\Phi }\left( \theta _{1}+2\pi i,..,\theta _{n}\right) =Q_{n}^{\Phi
}\left( \theta _{1},..,\theta _{n}\right)  \label{q3} \\
&&Q_{n+2}^{\Phi }\left( \theta +i\pi ,\theta ,\theta _{1},..,\theta
_{n}\right) =(-1)^{n}x\,D_{n}\left( x,x_{1},..,x_{n}\right) Q_{n}^{\Phi
}\left( \theta _{1},..,\theta _{n}\right) \,,  \label{q4}
\end{eqnarray}
where $x\equiv e^{\theta }$ and 
\begin{equation}
D_{n}\left( x,x_{1},..,x_{n}\right)
=\sum_{k=1}^{n}\sum_{m=1,odd}^{k}(-1)^{k+1}[m]\,x^{2(n-k)+m}\sigma
_{k}^{(n)}\sigma _{k-m}^{(n)}\,,  \label{Dn}
\end{equation}
with 
\begin{equation}
[m]\equiv\frac{\sin(m\xi)}{\sin\xi}\,.
\end{equation}

The equations (\ref{q1})-(\ref{q4}) admit infinitely many solutions which
account for the infinitely many operators with spin $s_{\Phi }$. The scalar (%
$s_{\Phi }=0$) solutions with the mildest asymptotic behavior are expected
to correspond to the primary operators of the theory, namely the exponential
operators (\ref{vertex}). Introducing the notation 
\begin{equation}
\hat{\Phi}=\frac{\Phi }{\langle \Phi \rangle }\,,
\end{equation}
the asymptotic factorization condition 
\begin{equation}
\lim_{\lambda \rightarrow +\infty }F_{n}^{\hat{\Phi}_{0}}\left( \theta
_{1}+\lambda ,\ldots ,\theta _{k}+\lambda ,\theta _{k+1},\ldots ,\theta
_{n}\right) =F_{k}^{\hat{\Phi}_{0}}(\theta _{1},\ldots ,\theta _{k})F_{n-k}^{%
\hat{\Phi}_{0}}\left( \theta _{k+1},\ldots ,\theta _{n}\right)
\label{cluster}
\end{equation}
characterizes the scalar primary operators $\Phi _{0}$ with non-vanishing
matrix elements on any number of particles \cite{DSC}, and in this case
selects the solutions \cite{KM} 
\begin{equation}
Q_{n}^{(a)}(\theta _{1},\ldots ,\theta _{n})=[a]\det M^{(n)}(a),  \label{Qna}
\end{equation}
where $M^{(n)}(a)$ is the $(n-1)\times (n-1)$ matrix with entries 
\begin{equation}
M_{i,j}^{(n)}(a)=[a+i-j]\,\sigma _{2i-j}^{(n)}
\end{equation}
and $a$ is a complex parameter. The trigonometric identity $%
[a]^{2}-[a-1][a+1]=1$ is useful to check that the functions (\ref{Qna})
solve the recursive equation (\ref{q4}). It was found in \cite{KM} (see also 
\cite{Luky}) that the solution (\ref{Qna}) corresponds to the exponential
operator\footnote{%
Since the identity $[a+2\pi /\xi ]=[a]$ implies the periodicity $%
Q_{n}^{(a+2\pi /\xi )}=Q_{n}^{(a)}$, we consider values of $a$ in the range $%
(0,2\pi /\xi )$.} $\hat{V}_{iab}=\hat{V}_{-a\beta }$. The solutions (\ref
{Qna}) satisfy the property 
\begin{equation}
Q_{n}^{(a)}=(-1)^{n}\,Q_{n}^{(-a)}\,,  \label{parity}
\end{equation}
which is expected since $a\rightarrow -a$ amounts to $\varphi \rightarrow
-\varphi $, and the particle $B$ is odd under this transformation. In
presence of the charge at infinity $Q=ib\pi /\xi $ we expect the
identification $\hat{V}_{\alpha }=\hat{V}_{Q-\alpha }$, namely 
\begin{equation}
Q_{n}^{(a)}=Q_{n}^{(-a-\pi /\xi )}\,\,.  \label{reflection}
\end{equation}
The identity 
\begin{equation}
\lbrack a]=-[a+\pi /\xi ]\,,
\end{equation}
together with (\ref{parity}), ensures that (\ref{reflection}) holds.

The solutions for the exponential operators allow to determine the form
factors of the components of the energy-momentum tensor. Let us start with
the trace (\ref{trace}). As seen in the previous section, when the
ultraviolet limit is taken to be the Liouville theory (\ref{AL}) we have%
\footnote{%
From now on we omit the tilde on the components of the energy-momentum
tensor.} ${\Theta }\sim V_{ib}$, so that 
\begin{equation}
Q_{n}^{{\Theta }}=-\frac{\pi \,m^{2}}{8\sin \xi }\,Q_{n}^{(1)}\,,
\label{Qtheta}
\end{equation}
where the normalization is fixed by the condition 
\begin{equation}
F_{2}^{\Theta }(\theta +i\pi ,\theta )=\frac{\pi }{2}\,m^{2}
\label{normtheta}
\end{equation}
which corresponds to the normalization 
\begin{equation}
\langle B(\theta )|B(\theta ^{\prime })\rangle =2\pi \,\delta (\theta
-\theta ^{\prime })
\end{equation}
of the asymptotic states. If instead the sinh-Gordon model is seen as a
perturbation of the Gaussian fixed point, the symmetric combination $%
Q_{n}^{(1)}+Q_{n}^{(-1)}$ must be considered. Then (\ref{parity}) shows that
the result (\ref{Qtheta}) still holds for $n$ even, while $Q_{n}^{\Theta }$
vanishes for $n$ odd. This analysis was first performed in \cite{MS}, where
the ultraviolet central charge was also evaluated through the $C$-theorem
sum rule \cite{cth,Cardycth} 
\begin{equation}
C=\frac{12}{\pi }\int d^{2}x\,|x|^{2}\,\langle \Theta (x)\Theta (0)\rangle
_{conn}\,\,.  \label{cth}
\end{equation}
It was checked that a truncated spectral expansion of the two-point trace
correlator in terms of the form factors (\ref{Qtheta}) reproduces with good
approximation the Liouville central charge (\ref{C}) if the sum is performed
over all $n$, and the Gaussian value $1$ if the sum is restricted to the
even contributions \cite{MS}.

When inserted in the asymptotic factorization equation (\ref{cluster}) with $%
\Phi=\Theta$ the solution (\ref{Qtheta}) prescribes the result 
\begin{equation}
\langle \Theta\rangle=F_{0}^{\Theta }=-\frac{\pi m^{2}}{8\sin\xi}\,,
\label{vevtheta}
\end{equation}
which coincides with that known from the thermodynamic Bethe ansatz (see 
\cite{AlyoshaSG}).

The form factors of the other components of the energy-momentum tensor are
easily obtained exploiting the conservation equations\footnote{%
We use the notation $\partial =\partial _{z}$ and $\bar{\partial}={\partial }%
_{\bar{z}}$ with reference to the complex coordinates $z=x_{1}+ix_{2}$ and $%
\bar{z}=x_{1}-ix_{2}$.} 
\begin{eqnarray}
&&\bar{\partial}T=\partial \Theta  \notag \\
&&\partial \bar{T}=\bar{\partial}\Theta
\end{eqnarray}
which lead to 
\begin{eqnarray}
&&F_{n}^{T}(\theta _{1},\ldots ,\theta _{n})=-\frac{\sigma _{1}^{(n)}\sigma
_{n}^{(n)}}{\sigma _{n-1}^{(n)}}\,F_{n}^{\Theta }(\theta _{1},\ldots ,\theta
_{n})  \notag \\
&&F_{n}^{\bar{T}}(\theta _{1},\ldots ,\theta _{n})=-\,\frac{\sigma
_{n-1}^{(n)}}{\sigma _{1}^{(n)}\sigma _{n}^{(n)}}F_{n}^{\Theta }(\theta
_{1},\ldots ,\theta _{n})  \label{FnT}
\end{eqnarray}
for $n>0$; $\langle T\rangle =\langle \bar{T}\rangle =0$ as for any operator
with nonzero spin.

It follows from the discussion of the previous section that, through the
analytic continuation (\ref{beta}), the above results for the exponential
operators in the sinh-Gordon model also hold for the matrix elements of
these operators on the breather $B_{1}$ of the sine-Gordon model. Moreover,
when the coupling $\xi $ takes the discrete values (\ref{xiN}) corresponding
to the reduction to the $\phi _{1,3}$-perturbed minimal models $\mathcal{M}%
_{2/(2N+3)}$, these results give, through the correspondence (\ref{phimn}),
the form factors for the independent primary operators $\phi _{1,l}$, $%
l=1,\ldots ,N+1$, of these massive minimal models. The reduction from the
continuous spectrum of exponential operators of the sinh-Gordon model to the
finite discrete spectrum of primary fields in the massive minimal models
follows from the fact that in the latter case the form factors have to
satisfy constraints on the bound state poles in addition to (\ref{fn0})-(\ref
{fn4}). For example, the fusion $B_{1}B_{1}\rightarrow B_{2}$ requires 
\begin{equation}
\mbox{Res}_{\theta ^{\prime }=\theta +i\xi }F_{n+2}^{\Phi }(\theta ^{\prime
},\theta ,\theta _{1},\ldots ,\theta _{n})=i\Gamma _{11}^{2}\langle 0|\Phi
(0)|B_{2}(\theta ^{\prime \prime })B_{1}(\theta _{1})\ldots B_{1}(\theta
_{n})\rangle \,,  \label{fusion}
\end{equation}
where $\theta ^{\prime \prime }=\theta +i\xi /2$ and the three-particle
coupling $\Gamma _{11}^{2}$ is obtained from 
\begin{equation}
\mbox{Res}_{\theta =i\xi }S_{11}(\theta )=i\left( \Gamma _{11}^{2}\right)
^{2}\,\,.
\end{equation}
It turns out \cite{Smirnovreduction,Koubek} that for $\xi =\xi _{N}$ the
complete set of bound state equations implied by the breather sector seen as
a self-contained bootstrap system selects among the solutions\footnote{%
We stress that for $\xi =\xi _{N}$ the knowledge of the $Q_{n}^{(a)}$ for
all $n$ completely determines the operator since all matrix elements
involving particles $B_{j}$ with $j>1$ can be obtained through fusion
equations like (\ref{fusion}).} (\ref{Qna}) only those with $a=1,\ldots ,N$,
besides the identity (see appendix \ref{Ap-M(2,2N+3)-boundstate}). These
solutions correspond to the operators 
\begin{equation}
\hat{V}_{-k\beta _{N}}=\hat{\phi}_{1,2k+1}\,,\hspace{1cm}k=1,\ldots ,N
\end{equation}
($\beta _{N}=-\sqrt{2/(2N+3)}$), which, in view of the reflection relation 
\begin{equation}
\hat{\phi}_{1,l}=\hat{\phi}_{1,2N+3-l}\,,\hspace{1cm}l=1,\ldots ,2N+2
\end{equation}
are all the primaries of this series of minimal models. The results (\ref
{Qtheta}), (\ref{vevtheta}) and (\ref{FnT}) for the matrix elements of the
energy-momentum tensor apply to the minimal massive models for $\xi =\xi
_{N} $.

\resection{The operator $T\bar{T}$} \label{Section-TTbar}We have seen in the
previous section how the primary operators correspond to the solutions of
the form factor equations (\ref{fn0})-(\ref{fn4}) with the mildest
asymptotic behavior at high energies. The remaining solutions of these
equations should span the space of descendant operators.

At criticality descendant operators are obtained acting on a primary with
products of Virasoro generators $L_{-i}$ and $\bar{L}_{-j}$. The sum of the
positive integers $i$ ($j$) defines the right (left) level $l$ ($\bar{l}$)
of the descendant. We denote by $\Phi_{l,\bar{l}}$ a descendant of level $(l,%
\bar{l})$ of a primary $\Phi_{0,0}\equiv\Phi_0$. Due to the isomorphism
between critical and off-critical operator spaces, the notion of level holds
also in the massive theory. A relation between the level and the asymptotic
behavior of form factors has been introduced in \cite{ttbar,seven}. In
particular, for operators $\Phi_{l,l}$ with non-zero matrix elements on any
number of particles this relation reads 
\begin{equation}
F_n^{\Phi_{l,l}}(\theta_1+\lambda,\ldots,\theta_k+\lambda,\theta_{k+1},
\ldots,\theta_n)\sim e^{l\lambda}\,,\hspace{1cm}\lambda\to +\infty
\label{asymp}
\end{equation}
for $n>1$ and $1\leq k\leq n-1$. It was also argued that the scaling
operators of this kind satisfy the asymptotic factorization property \cite
{ttbar} 
\begin{equation}
\lim_{\lambda\rightarrow +\infty}e^{-l\lambda}F_{n}^{\mathcal{L}_{l}\bar{%
\mathcal{L}}_{{l}}\Phi_0}\left(\theta_{1}+\lambda,\ldots,\theta_{k}+\lambda
,\theta _{k+1},\ldots,\theta_{n}\right)=\frac{1}{\left\langle\Phi_0
\right\rangle}F_{k}^{\mathcal{L}_{l}\Phi_0}\left(\theta_{1},\ldots,\theta
_{k}\right)F_{n-k}^{\bar{\mathcal{L}}_{{l}}\Phi_0}\left(\theta_{k+1},
\ldots,\theta_{n}\right),  \label{clusterdesc}
\end{equation}
where $\mathcal{L}_l$ and $\bar{\mathcal{L}}_l$ are operators that in the
conformal limit converge to the product of right and left Virasoro
generators, respectively, acting on the primary. This equation reduces to (%
\ref{cluster}) in the case of primary operators.

This asymptotic information can be used to classify the form factor
solutions according to the level and to determine the leading part in the
high-energy (conformal) limit. The subleading contributions depend instead
on the way the operators are defined off-criticality. A complete analysis of
the operator space in the massive Lee-Yang model up to level 7 is given in 
\cite{seven}.

The composite operator $T\bar{T}$, obtained from the non-scalar components
of the energy-momentum tensor, is the simplest non-derivative scalar
descendant of the identity. A.~Zamolodchikov showed that this operator can
be defined away from criticality as \cite{Sasha} 
\begin{equation}
T\bar{T}(x)=\lim_{\epsilon \rightarrow 0}[T(x+\epsilon )\bar{T}(x)-\Theta
(x+\epsilon )\Theta (x)+\mbox{derivative terms}]\,\text{,}
\label{regularized}
\end{equation}
where 'derivative terms' means terms containing powers of $\epsilon $ times
local operators which are total derivatives. One consequence of this
equation is that, if $|n\rangle $ and $|m\rangle $ denote $n$- and $m$%
-particle states with the same energy ($E_{n}=E_{m}$) and momentum ($%
P_{n}=P_{m}$), the equation 
\begin{equation}
\langle m|T\bar{T}(0)|n\rangle =\langle m|T(x)\bar{T}(0)|n\rangle -\langle
m|\Theta (x)\Theta (0)|n\rangle  \label{TTmn}
\end{equation}
holds, with the r.h.s. that does not depend on $x$. Since the generic matrix
element can be reduced to the form factors (\ref{ff}) iterating the crossing
relation 
\begin{align}
\langle B(\theta _{m}^{\prime })\ldots B(\theta _{1}^{\prime })|\Phi
(0)|B(\theta _{1})\ldots B(\theta _{n})\rangle =\langle B(\theta
_{m}^{\prime })\ldots B(\theta _{2}^{\prime })|\Phi (0)|B(\theta
_{1}^{\prime }+i\pi )B(\theta _{1})\ldots B(\theta _{n})\rangle +&  \notag \\
2\pi \sum_{i=1}^{n}\delta (\theta _{1}^{\prime }-\theta
_{i})\prod_{k=1}^{i-1}S(\theta _{k}-\theta _{1}^{\prime })\,\langle B(\theta
_{m}^{\prime })\ldots B(\theta _{2}^{\prime })|\Phi (0)|B(\theta _{1})\ldots
B(\theta _{i-1})B(\theta _{i+1})\ldots B(\theta _{n})\rangle \,& ,
\label{crossing}
\end{align}
the identities (\ref{TTmn}) contribute to the identification of the form
factor solution for the operator $T\bar{T}$, in particular of the subleading
parts which are left unconstrained by the asymptotic factorization property (%
\ref{clusterdesc}). Since $T\bar{T}$ is a level $(2,2)$ descendant of the
identity, the factorization takes in this case the form 
\begin{equation}
\lim_{\lambda \rightarrow +\infty }e^{-2\lambda }F_{n}^{T\bar{T}}\left(
\theta _{1}+\lambda ,\ldots ,\theta _{k}+\lambda ,\theta _{k+1},\ldots
,\theta _{n}\right) =F_{k}^{T}\left( \theta _{1},\ldots ,\theta _{k}\right)
F_{n-k}^{\bar{T}}\left( \theta _{k+1},\ldots ,\theta _{n}\right) \,.
\label{clusterttbar}
\end{equation}

We now argue that (\ref{TTmn}) implies the property 
\begin{equation}
F_n^{T\bar{T}}=\mbox{sum of terms containing}\,\,\,(\sigma_1^{(n)})^i(%
\sigma_{n-1}^{(n)})^j\,,\hspace{1cm}i+j\geq 2  \label{factor}
\end{equation}
for $n$ larger than 2 but other than 4. Observe first that the components of
the energy-momentum tensor, being local operators of the theory, must have
form factors whose only singularities in rapidity space are the annihilation
poles prescribed by (\ref{fn4}), and possible bound state poles \cite
{AlyoshaYL}. Then, it follows from (\ref{FnT}) that 
\begin{equation}
F_n^{T}\propto(\sigma_1^{(n)})^2\,,\hspace{1cm} F_n^{\bar{T}%
}\propto(\sigma_{n-1}^{(n)})^2\,,\hspace{1cm} F_n^{\Theta}\propto%
\sigma_1^{(n)}\sigma_{n-1}^{(n)}\,,  \label{components}
\end{equation}
for $n>2$. Use now the resolution of the identity 
\begin{equation}
I=\sum_{k=0}^{\infty }\frac{1}{k!}\int \frac{d\theta _{1}}{2\pi }\cdots 
\frac{d\theta _{k}}{2\pi }\,|k\rangle \langle k|  \label{identity}
\end{equation}
to expand the r.h.s. of (\ref{TTmn}) over matrix elements of $T$, $\bar{T}$
and $\Theta$. If the total energy-momentum of the intermediate state $%
|k\rangle$ differs from that of $|m\rangle$ (which, we recall, equals that
of $|n\rangle$) the two sums in the r.h.s. separately depend on $x$, and
must cancel each other in order to ensure the $x$-independence of the
result. Then we are left with the contributions of matrix elements over
states with the same energy and momentum, which all are $x$-independent.
Consider the case $m$ and $n$ both larger than zero, $m\neq n$. It follows
from (\ref{components}) that each of these matrix elements generically
vanishes at least as $\eta^2$, if $\eta$ is an infinitesimal splitting
between the energies of the two states in the matrix element. A milder
behavior as $\eta\to 0$ is obtained when $|k\rangle$ is identical to $%
|m\rangle$ or to $|n\rangle$. Indeed it can be shown using (\ref{crossing})
and (\ref{fn4}) that the matrix elements of the energy-momentum tensor over
identical states are finite and non-zero. In the r.h.s. of (\ref{TTmn})
these non-zero matrix elements multiply a matrix element vanishing at least
as $\eta^2$. We argue in a moment that the form of the `derivative terms' in
(\ref{regularized}) is such that they contribute terms vanishing at least as 
$\eta^2$ to the r.h.s. of (\ref{TTmn}). Putting all together, we conclude
that the l.h.s. vanishes at least as $\eta^2$, and this implies the form (%
\ref{factor}). A more detailed derivation including the explanation of the
limitations on $n$ can be found in appendix~B together with the form factor
expansion of (\ref{TTmn}).

The property (\ref{factor}) can also be understood in the following way.
Since $m\sigma_1^{(n)}$ and $m\sigma_{n-1}^{(n)}/\sigma_n^{(n)}$ are the
eigenvalues of $P=i\partial$ and $\bar{P}=-i\bar{\partial}$ on an $n$%
-particle asymptotic state, (\ref{components}) follows from the fact that in
two dimensions the energy-momentum tensor can formally be written as $%
T_{\mu\nu}(x)=(2/\pi)(\eta_{\mu\nu}\Box-\partial_\mu\partial_\nu)A(x)$, or 
\begin{equation}
T=\partial^2 A\,,\hspace{1cm}\bar{T}=\bar{\partial}^2 A\,,\hspace{1cm}%
\Theta=\partial\bar{\partial}A\,,
\end{equation}
in terms of an operator $A(x)$ which is \emph{not} a local operator of the
theory\footnote{%
Indeed, $F_2^A(\theta_1,\theta_2)$ contains a pole at $\theta_1-\theta_2=i%
\pi $, in contrast with (\ref{fn4}) which prescribes a vanishing residue.
Essentially, this is why (\ref{components}) holds only for $n>2$.}. Using
the notation $A\cdot B\equiv A(x+\epsilon)B(x)$ one has 
\begin{equation}
T\cdot\bar{T}-\Theta\cdot\Theta=\frac12\partial^2(A\cdot\bar{T})+\frac12\bar{%
\partial}^2(A\cdot T)- \partial\bar{\partial}(A\cdot\Theta)\,\,.  \label{dot}
\end{equation}
The property (\ref{factor}) then follows from the fact that also the
`derivative terms' in (\ref{regularized}) must be derivative operators of at
least second order: those associated to negative powers of $\varepsilon$
because they must cancel the divergences arising in (\ref{dot}) when $%
\epsilon\to 0$; those associated to $\epsilon^0$ because they must have $l=%
\bar{l}\leq 2$ and we know that $L_{-1}=\partial$, $\bar{L}_{-1}=\bar{%
\partial}$. %The properties 
%(\ref{components}) and (\ref{factor}) hold in two dimensions even in absence of integrability.

Having collected this information, we can move forward in the determination
of the form factors of $T\bar{T}$. The first requirement to be satisfied is
that they solve the form factor equations (\ref{fn0})-(\ref{fn4}) and have
the asymptotic behavior (\ref{asymp}) with $l=2$. Since the equations (\ref
{fn0})-(\ref{fn4}) are linear in the operator, the solution for $T\bar{T}$
can be written as a linear superposition of the scalar form factor solutions
behaving as in (\ref{asymp}) with $l=0,1,2$. We will refer to these
solutions as level $0$, level $1$ and level $2$ solutions, respectively. We
now argue that this linear combination can be restricted to 
\begin{equation}
F_{n}^{T\bar{T}}=a\,m^{-2}\,F_{n}^{\partial ^{2}\bar{\partial}^{2}\Theta
}+F_{n}^{\mathcal{K}}+c\,F_{n}^{\partial \bar{\partial}\Theta
}+d\,m^{2}\,F_{n}^{\Theta }+e\,m^{4}\,F_{n}^{I}\,,  \label{fnttbar}
\end{equation}
with $a$, $c$, $d$, $e$ dimensionless constants and $F_{n}^{\mathcal{K}}$ to
be defined below. Indeed, the level $0$ solutions corresponding to scaling
operators are spanned by the primaries and, among these, only the form
factors 
\begin{equation}
F_{n}^{I}=\delta _{n,0}
\end{equation}
of the identity and those of $\Theta $ satisfy the requirement (\ref{factor}%
). As for the level $1$ scaling operators, they are all of the form $L_{-1}%
\bar{L}_{-1}\Phi _{0}=\partial \bar{\partial}\Phi _{0}$. Equations (\ref
{TTmn}), (\ref{clusterttbar}) and (\ref{factor}) put no constraint on the
contribution to (\ref{fnttbar}) of such level 1 derivative operators.
However, we expect that the form factors of the operator $T\bar{T}$ in the
sinh-Gordon model enjoy the following properties\footnote{%
Analogous properties do hold for the components $T$, $\bar{T}$ and $\Theta $
of the energy-momentum tensor.}:

i) to give the form factors of $T\bar{T}$ on the lightest breather of the
sine-Gordon model under analytic continuation to positive values of $\xi$;

ii) to give the form factors of $T\bar{T}$ on the lightest particle of the $%
\phi_{1,3}$-perturbed minimal models $\mathcal{M}_{2/(2N+3)}$ when we set $%
\xi$ to the values $\xi_N$ defined by (\ref{xiN});

iii) to be continuous functions of $\xi $.

\noindent We know that, while the sinh-Gordon and sine-Gordon models possess
a continuous spectrum of primary operators, the $\phi _{1,3}$-perturbed
minimal models $\mathcal{M}_{2/(2N+3)}$ possess a discrete spectrum of $N+1$
primaries. The identity and the trace of the energy-momentum tensor are the
only primary operators which are present in all these models. Then the
natural way to comply with the requirements i)-iii) is that $I$ and $\Theta $
are the only primaries that contribute to (\ref{fnttbar}). The extension of
the argument to any of the models (\ref{Arsg}) with $p/p^{\prime }$ rational
implies that this is actually the only possibility. While we had reached
this conclusion about the primaries by another path, the present reasoning
also requires that $\partial \bar{\partial}\Theta $ and $\partial ^{2}\bar{%
\partial}^{2}\Theta $ are the only derivative operators which can appear in (%
\ref{fnttbar}).

The superposition (\ref{fnttbar}) without the term $F_{n}^{\mathcal{K}}$ is
sufficient to provide the most general parameterization with the required
asymptotic behavior up to $n=2$. This is why $F_{n}^{\mathcal{K}}$ is a
scalar three-particle kernel solution of the form factor equations, namely
has the property 
\begin{equation}
F_{n}^{\mathcal{K}}=0\,,\hspace{1cm}n=0,1,2\,;  \label{k3}
\end{equation}
the first two non-vanishing elements of this solution are $F_{3}^{\mathcal{K}%
}$ and $F_{4}^{\mathcal{K}}$, with $U_{3}^{\mathcal{K}}$ and $U_{4}^{%
\mathcal{K}}$ which factorize $\prod_{i<j}\cosh\frac{\theta_i-\theta_j}{2}$
in such a way to satisfy (\ref{fn4}) with $0$ on the r.h.s. The $F_{n}^{%
\mathcal{K}}$ are made of terms which under the limit (\ref{asymp}) behave
as $e^{l\lambda}$ with $l=0,1,2$. They define the local operator ${\mathcal{K%
}}(x)$ %\begin{equation}
%{\mathcal{K}}(x)=T\bar{T}(x)-a\,m^{-2}\,{\partial^{2}\bar{\partial}^{2}\Theta}(x)-
%c\,\partial\bar{\partial}\Theta(x)-d\,m^{2}\,{\Theta}(x)-e\,m^{4}\,I\,,
%\label{kernel}
%\end{equation}
which in (\ref{fnttbar}) accounts for the linear independence of $T\bar{T}$
within the operator space of the theory.

Since our equations do not constrain the level $1$ derivative contributions
to (\ref{fnttbar}), the coefficient $c$ will remain undetermined. This
conclusion agrees with conformal pertubation theory, which states that, due
to the resonance phenomenon \cite{Taniguchi} (see also \cite{FFLZZ}), the
operator $T\bar{T}$ can only be defined up to a term proportional to $%
\partial \bar{\partial}\Theta (x)$ \cite{Sasha}. Our remaining task is that
of showing that the r.h.s. of (\ref{fnttbar}) can be uniquely determined up
to this ambiguity.

For $n=2$, the second derivative term is the only one contributing to the
limit in (\ref{clusterttbar}), and this fixes 
\begin{equation}
a=\frac{\langle \Theta \rangle }{m^{2}}\,\,.  \label{a}
\end{equation}
The coefficients $d$ and $e$ can be determined by (\ref{TTmn}) with $m=n=1$.
Indeed, when we use (\ref{identity}) to expand the operator product in the
r.h.s., the $x$-independence of the result implies that only the $k=1$
intermediate state gives a non-vanishing contribution. Using (\ref{crossing}%
) to go to form factors we obtain the identities\footnote{%
The identity (\ref{TT00}) was originally observed in \cite{FFLZZ} and
follows also from (\ref{TTmn}) with $m=n=0$ and $|x|\rightarrow \infty $.
See also \cite{BS} for results on the vacuum expectation values of
descendant operators in integrable models.} 
\begin{equation}
F_{2}^{T\bar{T}}(i\pi ,0)=-2\langle \Theta \rangle F_{2}^{\Theta }(i\pi
,0)=-\pi m^{2}\langle \Theta \rangle  \label{TT11}
\end{equation}
\begin{equation}
\langle T\bar{T}\rangle =-\langle \Theta \rangle ^{2}\,\,.  \label{TT00}
\end{equation}
On the other hand, we have from (\ref{fnttbar}) that 
\begin{equation}
\langle T\bar{T}\rangle =d\,m^{2}\,\langle \Theta \rangle +e\,m^{4}
\end{equation}
\begin{equation}
F_{2}^{T\bar{T}}(i\pi ,0)=d\,m^{2}\,F_{2}^{\Theta }(i\pi ,0)\,,
\end{equation}
so that we obtain 
\begin{equation}
d=-\frac{2}{m^{2}}\,\langle \Theta \rangle
\end{equation}
\begin{equation}
e=\frac{\langle \Theta \rangle ^{2}}{m^{4}}\,.
\end{equation}

The search for the kernel contribution $F_{n}^{\mathcal{K}}$ to (\ref
{fnttbar}) starts from the most general solution of the form factor
equations satisfying (\ref{k3}), (\ref{factor}) and (\ref{asymp}) with $l=2$%
. We checked that equations (\ref{clusterttbar}) and (\ref{TTmn}) uniquely
fix the level $2$ and level $0$ parts, respectively, within such a solution.
Concerning the level $1$ part, it cannot be determined by these conditions,
because they do not exclude the contribution of those linear combinations of 
$\partial \bar{\partial}$-derivatives of the primaries which vanish on one-
and two-particle states. This indetermination is eliminated if we impose the
conditions i)-iii) above. To see this consider the $\phi _{1,3}$-perturbed
minimal models $\mathcal{M}_{2/(2N+3)}$. The identification $%
B_{2N+1-n}\equiv B_{n}$, $n=1,\dots,N$, that follows from (\ref{periodicity}%
), implies the set of equations 
\begin{equation}
\langle 0|\Phi (0)|B_{2N+1-n}\rangle =\langle 0|\Phi (0)|B_{n}\rangle
\label{B(2N+1-n)=B(n)}
\end{equation}
for any local operator $\Phi$ of the massive minimal model. In the minimal
models, and more generally in the sine-Gordon model, the matrix elements
involving particles $B_{n}$ with $n>1$ are related to the form factors (\ref
{ff}) by residue equations on bound state poles of the type (\ref{fusion}).
Due to (\ref{periodicity}), in the minimal models these bound state
equations constrain the form factors (\ref{ff}) themselves. For example, in
the simplest case, $N=1$, equation (\ref{fusion}) holds with $B_2$
identified to $B_1$ (see appendix~E for the case of generic $N$). We find
that the only way of satisfying (\ref{B(2N+1-n)=B(n)}) for any $N$ when $%
\Phi=T\bar{T}$ is to take\footnote{%
For $n=1,2$ (\ref{oneparticle}) is implied by (\ref{k3}) for any value of $%
\xi$.} 
\begin{equation}
\langle 0|\mathcal{K}(0)|B_{n}\rangle =0  \label{oneparticle}
\end{equation}
for $n=1,\ldots ,2N$, or equivalently\footnote{%
The equivalence between (\ref{oneparticle}) and (\ref{boundstate}) for $n>2N$
follows from the periodicity at $\xi =\xi _{N}$.} 
\begin{equation}
\lim_{\eta \rightarrow 0}\eta ^{n-1}F_{n}^{\mathcal{K}}\left( \theta
_{1}+\eta ,\theta _{2}+2\eta ,\ldots ,\theta _{n}+n\eta \right) =0
\label{boundstate}
\end{equation}
with $\theta _{k}=\theta _{1}+i(k-1)\xi $, $k=2,\ldots,n$ and $n\geq 1$.

The requirement of continuity in $\xi $ then leads to extend (\ref
{boundstate}) to generic values of $\xi $. We checked explicitly up to $n=9$
that the conditions (\ref{TTmn}), (\ref{clusterttbar}) and (\ref{boundstate}%
) uniquely determine $F_{n}^{\mathcal{K}}$ for generic $\xi $. The explicit
derivation up to $n=4$ is given in appendix~C, while appendix~D contains the
list of results up to $n=7$. When inserted in (\ref{fnttbar}) they provide
the form factors of $T\bar{T}$ in the sinh-Gordon model seen as a
perturbation of Liouville theory. The results apply to the sine-Gordon and
the $\phi_{1,3}$-perturbed minimal models $\mathcal{M}_{2/(2N+3)}$ as
specified by i) and ii) above. Setting to zero the $F_n^{T\bar{T}}$ with $n$
odd one obtains the result for the sinh-Gordon and sine-Gordon models seen
as perturbations of the Gaussian fixed point, for which the reflection
symmetry $\varphi\to-\varphi$ holds. Few remarks about the free limit $%
\xi\to 0$ are contained in appendix~F.

\resection{Conclusion} In this paper we identified the form factor solution
corresponding to the operator $T\bar{T}$ in the sinh-Gordon model and in the 
$\phi_{1,3}$-perturbed minimal models $\mathcal{M}_{2/(2N+3)}$. The
identification is obtained up to the arbitrary additive contribution of the
operator $\partial\bar{ \partial}\Theta$ which represents an intrinsic
ambiguity in the definition of $T\bar{T}$.

We expect that the possibility of expressing the solution for $T\bar{T}$ as
the superposition (\ref{fnttbar}) of a three-particle kernel solution plus
the contributions of the identity and the trace of the energy-momentum
tensor together with its first two scalar derivatives is not specific to the
class of models considered in this paper but is actually quite general. If
so, the expressions of the coefficients $a$, $d$ and $e$ given here should
be universal.

The property, expressed by (\ref{dot}) and (\ref{factor}), that $T\bar{T}$
behaves as a linear combination of derivative operators on states with more
than four particles should also be very general. In particular, we remarked
in \cite{ttbar} that it ensures that $T\bar{T}$ does not contribute to
integrability breaking (at least to first order) when used to perturb a
fixed point action. As a matter of fact, many examples are known of
integrable massless flows in which $T\bar{T}$ is the leading operator
driving the flow into the infrared fixed point \cite{tim}.

\vspace{1cm} \textbf{Acknowledgments.}~~This work was partially supported by
the European Commission TMR programme HPRN-CT-2002-00325 (EUCLID) and by the
MIUR programme ``Quantum field theory and statistical mechanics in low
dimensions''. G.N. was supported by a postdoctoral fellowship of the
Minist\`{e}re fran\c{c}ais d\'{e}l\'{e}gu\'{e} \`{a} l'Enseignement
sup\'{e}rieur et \`{a} la Recherche.

\appendix
\resection{Appendix} Consider the free massive boson described by the action
(\ref{A}) with 
\begin{equation}
\mathcal{V}=-\frac{1}{2}m^{2}\varphi ^{2}\,.
\end{equation}
The local scalar operators of the theory satisfy (\ref{fn0})-(\ref{fn4})
with $s_{\Phi }=0$ and $S(\theta )=1$. The primary operators 
\begin{equation}
V_{\alpha }=e^{i\sqrt{8\pi }\,\alpha \,\varphi }=\sum_{n=0}^{\infty }\frac{1%
}{n!}\,(i\sqrt{8\pi }\,\alpha \,\varphi )^{n}
\end{equation}
are subject also to the factorization property (\ref{cluster}). One then
obtains 
\begin{equation}
F_{n}^{V_{\alpha }}=\left( i\sqrt{8\pi }\,\alpha \,F_{1}^{\varphi }\right)
^{n}
\end{equation}
\begin{equation}
F_{n}^{\varphi ^{n}}=n!\,\left( F_{1}^{\varphi }\right) ^{n}\,\,.
\end{equation}
Equation (\ref{thetacl}) gives in this free case\footnote{%
We write $\Theta $ for $\tilde{\Theta}$.} 
\begin{equation}
\Theta =\Theta _{cl}=\frac{\pi m^{2}}{2}\left( \varphi ^{2}-\frac{iQ}{\sqrt{%
2\pi }}\,\varphi \right) \,,
\end{equation}
and then 
\begin{eqnarray}
&&F_{1}^{\Theta }=-\frac{m^{2}}{2}\sqrt{\frac{\pi }{2}}\,iQ\,F_{1}^{\varphi }
\\
&&F_{2}^{\Theta }=\pi m^{2}\,\left( F_{1}^{\varphi }\right) ^{2}
\end{eqnarray}
as the only non-zero form factors of this operator. Comparison with (\ref
{normtheta}) fixes 
\begin{equation}
F_{1}^{\varphi }=\frac{1}{\sqrt{2}}\,\,.
\end{equation}
The results (\ref{C}) and (\ref{X}) then easily follow using (\ref{identity}%
) to evaluate over form factors the sum rules (\ref{cth}) and \cite{DSC} 
\begin{equation}
X_{\Phi }=-\frac{2}{\pi \langle \Phi \rangle }\int d^{2}x\,\langle \Theta
(x)\Phi (0)\rangle _{conn}\,\,.
\end{equation}

\resection{Appendix} \label{Ap-vev-development}Here we derive the
constraints imposed on the form factors of $T\bar{T}$ by the relations (\ref
{TTmn}) and show how they lead to the property (\ref{factor}).

The r.h.s. of (\ref{TTmn}) is expanded introducing in between the two pairs
of operators the resolution of the identity (\ref{identity}). Then the l.h.s
and r.h.s. of (\ref{TTmn}) are rewritten in terms of form factors by
iterative use of the crossing relation (\ref{crossing}). Let $\left\langle
m\right| =\langle B(\theta _{m}^{\prime })\ldots B(\theta _{1}^{\prime })|$\
and $|n\rangle =|B(\theta _{1})\ldots B(\theta _{n})\rangle $ be the two
states with the same energy and momentum in (\ref{TTmn}).

In order to avoid to sit directly on annihilation or bound state poles of
the form factors we introduce an infinitesimal splitting, parametrized by $%
\eta $, between the energies and momenta of these two states. The identity (%
\ref{TTmn}) is recovered in the limit $\eta \rightarrow 0$.

The l.h.s. has an expansion of the form: 
\begin{align}
& \left. \langle B(\theta _{m}^{\prime }+\eta _{m}^{\prime })\ldots B(\theta
_{1}^{\prime }+\eta _{1}^{\prime })|T\bar{T}(0)|B(\theta _{1}+\eta
_{1})\ldots B(\theta _{n}+\eta _{n})\rangle =\right.  \notag \\
& F_{n+m}^{T\bar{T}}\text{(}\theta _{m}^{\prime }\text{+}\eta _{m}^{\prime }%
\text{+}i\pi \text{,..,}\theta _{1}^{\prime }\text{+}\eta _{1}^{\prime }%
\text{+}i\pi \text{,}\theta _{1}\text{+}\eta _{1}\text{,..,}\theta _{n}\text{%
+}\eta _{n}\text{)}+2\pi \sum_{j=1}^{m}\sum_{i=1}^{n}\delta (\eta
_{j}^{\prime }-\eta _{i}+\theta _{j}^{\prime }-\theta _{i})  \notag \\
& \prod_{h=1}^{j-1}\prod_{k=1}^{i-1}S(\eta _{j}^{\prime }-\eta _{h}^{\prime
}+\theta _{j}^{\prime }-\theta _{h}^{\prime })\,S(\eta _{k}-\eta
_{j}^{\prime }+\theta _{k}-\theta _{j}^{\prime })\,F_{n+m-2}^{T\bar{T}}(\hat{%
\imath},\hat{\jmath})+...  \label{lhs-expansion}
\end{align}
where $\eta _{a}^{\prime }=(n+a)\eta ,$ $a=1,\ldots ,m,$\ $\eta _{b}=b\eta ,$
$b=1,\ldots ,n$, and $F_{n+m-2}^{T\bar{T}}(\hat{\imath},\hat{\jmath})$ is
the form factor with $n+m-2$ particles obtained omitting the particles $%
B(\theta _{i}+\eta _{i})$ and $B(\theta _{j}^{\prime }+\eta _{j}^{\prime })$%
. The dots in (\ref{lhs-expansion}) represent terms which factorize\ $p$
delta functions, a product of two-particle amplitudes, and form factors of $T%
\bar{T}$ with $n+m-2p$ particles with $p=2,...,$ Int$((n+m)/2)$.

Let us consider now the r.h.s. of (\ref{TTmn}). Among the infinitely many
terms generated by the insertion of (\ref{identity}) the only ones relevant
for the identity (\ref{TTmn}) are those that do not depend on $x$ for $\eta
=0$. For each fixed $k$, $\left\langle m\right| T(x)|k\rangle \left\langle
k\right| \bar{T}(0)|n\rangle -\left\langle m\right| \Theta (x)|k\rangle
\left\langle k\right| \Theta (0)|n\rangle $ can be expanded in terms of the
form factors by using expansions like (\ref{lhs-expansion}) for each matrix
element. Here the arguments of the delta functions are the differences
between the rapidities of the states $|m\rangle $ or $|n\rangle $ and those
of $|k\rangle $. In these expansions the terms that do not depend on $x$ are
only those factorizing a set of delta functions which saturate all the
integrations in (\ref{identity}) and fix the state $|k\rangle $ to one with
the same energy and momentum of $|n\rangle $ and $|m\rangle $. Such terms
are in finite number and are generated only if $k=n$ or $k=m$, when the
delta functions fix the state $|k\rangle $ to $|n\rangle $ or $|m\rangle $,
respectively. We can now rearrange such terms according to their content in
delta functions: we have a number of terms which do not factorize delta
functions plus a number of terms which factorize one delta function, and so
on as in formula (\ref{lhs-expansion}). The identity (\ref{TTmn}) is
recovered imposing that in the l.h.s. and in the r.h.s. the terms that
factorize the same delta functions coincide in the limit $\eta \rightarrow 0$%
. An iterative structure appears for these identities (see (\ref{TT11}) for
the particular case $m=n=1$). In general, the terms factorizing delta
functions give the results of the identities (\ref{TTmn}) with $m-1$ and $%
n-1 $, while the terms without delta functions give the new condition 
\begin{gather}
\lim_{\eta \rightarrow 0}F_{n+m}^{T\bar{T}}\text{(}\theta _{m}^{\prime }%
\text{+}\eta _{m}^{\prime }\text{+}i\pi \text{,..,}\theta _{1}^{\prime }%
\text{+}\eta _{1}^{\prime }\text{+}i\pi \text{,}\theta _{1}\text{+}\eta _{1}%
\text{,..,}\theta _{n}\text{+}\eta _{n}\text{)}=\lim_{\eta \rightarrow
0}\{\sum_{a=0}^{\min (n,m)}\frac{1}{2(n-a)!a!}\sum_{\tau \in \mathcal{P}%
_{n}}\sum_{\mu \in \mathcal{P}_{m}}\text{ \ \ \ }  \notag \\
\prod_{h=1}^{a}\Delta (\theta _{\mu _{m-a+h}}^{\prime }-\theta _{\tau
_{n-a+h}})\{[F_{m+n-2a}^{T}(\theta _{\mu _{m-a}}^{\prime }\text{+}\eta _{\mu
_{m-a}}^{\prime }\text{+}i\pi \text{,..,}\theta _{\mu _{1}}^{\prime }\text{+}%
\eta _{\mu _{1}}^{\prime }\text{+}i\pi \text{,}\theta _{\tau _{1}}\text{+}%
\eta _{\tau _{1}}\text{,..,}\theta _{\tau _{n-a}}\text{+}\eta _{\tau
_{n-a}})\times  \notag \\
F_{2a}^{\bar{T}}(\theta _{\mu _{m}}^{\prime }\text{+}\eta _{\mu
_{m}}^{\prime }\text{+}i\pi \text{,..,}\theta _{\mu _{m-a+1}}^{\prime }\text{%
+}\eta _{\mu _{m-a+1}}^{\prime }\text{+}i\pi \text{,}\theta _{\tau _{n-a+1}}%
\text{+}\eta _{\tau _{n-a+1}}\text{,..,}\theta _{\tau _{n}}\text{+}\eta
_{\tau _{n}})-\text{ \ \ \ \ \ \ \ \ \ \ \ \ \ \ \ \ \ \ \ \ \ \ \ \ \ \ \ \
\ \ }  \notag \\
F_{m+n-2a}^{\Theta }(\theta _{\mu _{m-a}}^{\prime }\text{+}\eta _{\mu
_{m-a}}^{\prime }\text{+}i\pi \text{,..,}\theta _{\mu _{1}}^{\prime }\text{+}%
\eta _{\mu _{1}}^{\prime }\text{+}i\pi \text{,}\theta _{\tau _{1}}\text{+}%
\eta _{\tau _{1}}\text{,..,}\theta _{\tau _{n-a}}\text{+}\eta _{\tau
_{n-a}})\times \text{ \ \ \ \ \ \ \ \ \ \ \ \ \ \ \ \ \ \ \ \ \ \ \ \ \ \ \
\ \ \ \ \ \ \ }  \notag \\
F_{2a}^{\Theta }(\theta _{\mu _{m}}^{\prime }\text{+}\eta _{\mu
_{m}}^{\prime }\text{+}i\pi \text{,..,}\theta _{\mu _{m-a+1}}^{\prime }\text{%
+}\eta _{\mu _{m-a+1}}^{\prime }\text{+}i\pi \text{,}\theta _{\tau _{n-a+1}}%
\text{+}\eta _{\tau _{n-a+1}}\text{,..,}\theta _{\tau _{n}}\text{+}\eta
_{\tau _{n}})]\text{+}[T\leftrightarrow \bar{T}]\}\text{ ,\ \ \ \ \ \ \ \ \
\ \ \ \ }  \label{F(n+m)-vev-general}
\end{gather}
where $\mathcal{P}_{n}$ is the group of permutations of $n$ indices, $\tau $
is a permutation and 
\begin{equation*}
\Delta (x)=\left\{ 
\begin{array}{c}
1\text{ \ for }x=0\text{ ,} \\ 
0\text{ \ for }x\neq 0\text{ .}
\end{array}
\right.
\end{equation*}
Let us observe now that the r.h.s. of (\ref{F(n+m)-vev-general}) goes always
to zero at least as $\eta ^{2}$, because the $p$-particle form factors of $T,%
\bar{T}$ and $\Theta $ respectively factorize $(\sigma _{1}^{(p)})^{2}$, $%
(\sigma _{p-1}^{(p)})^{2}$ and $\sigma _{1}^{(p)}\sigma _{p-1}^{(p)}$. The
only exceptions to this situation arise if some rapidities are grouped in
pairs that lie on annihilation poles or for rapidity configurations which
intercept bound state poles. In every case, in terms of the parametrization (%
\ref{fn}), the only way in which the form factors of $T\bar{T}$ can satisfy (%
\ref{F(n+m)-vev-general}) is that for any $p>2$ the function $U_{p}^{T\bar{T}%
}$ is the sum of terms each one factorizing $\left( \sigma _{1}^{(p)}\right)
^{i}\left( \sigma _{p-1}^{(p)}\right) ^{j}$ with some $i$ and $j$ such that $%
i+j\geq 2.$

This property amounts to (\ref{factor}) provided that there are no
cancellations with the denominator of (\ref{fn}). This denominator can be
written as 
\begin{eqnarray}
\prod_{1\leq i<j\leq p}\cosh(\theta_i-\theta_j) &=& \left(\frac{1}{%
2^p\sigma_p^{(p)}}\right)^{(p-1)/2}\prod_{1\leq i<j\leq p}(x_i+x_j)  \notag
\\
&=& \left(\frac{1}{2^p\sigma_p^{(p)}}\right)^{(p-1)/2}\mbox{det}\,D^{(p)}\,,
\label{deno}
\end{eqnarray}
where $D^{(p)}$ is the $(p-1)\times(p-1)$ matrix with entries 
\begin{equation}
D^{(p)}_{ij}=\sigma^{(p)}_{2i-j}\,\,.
\end{equation}
For $p>2$, it is only for $p=4$ that (\ref{deno}) is a sum of terms all
containing non-vanishing powers of $\sigma_1^{(p)}$ or $\sigma_{p-1}^{(p)}$
and cancellations may occur.

\resection{Appendix} \label{Ap-TTbar-up to 4} In this appendix we determine
the form factors of the kernel $\mathcal{K}$ up to four particles. At three
particles the most general solution of the form factor equations satisfying (%
\ref{factor}), (\ref{asymp}) with $l=2$ and the initial condition 
\begin{equation}
F_{1}^{\mathcal{K}}=0  \label{K1=0}
\end{equation}
is\footnote{%
We simplify the notation by dropping the superscript $(n)$ on the symmetric
polynomials.} 
\begin{equation}
\frac{b_{3}\sigma _{1}^{2}\sigma _{2}^{2}+\text{D}_{3}\sigma _{2}^{3}+\text{C%
}_{3}\sigma _{1}^{3}\sigma _{3}}{\sigma _{3}^{2}}F_{3}^{K_{3}}+\text{B}_{3}\,%
\frac{\sigma _{1}\sigma _{2}}{\sigma _{3}}F_{3}^{K_{3}}\text{ .}
\label{cin-sol-K3}
\end{equation}
Here, $F_{3}^{K_{3}}$ is defined by 
\begin{equation}
Q_{3}^{K_{3}}=(\sigma _{1}\sigma _{2}-\sigma _{3})\text{ }
\end{equation}
in the parametrization (\ref{fn}) and generates the one dimensional space of
the level 0 solutions of the equations (\ref{fn0})-(\ref{fn4}) which vanish
at one particle. The first and second term in (\ref{cin-sol-K3}) are pure
level 2 and level 1 parts, respectively, and $b_{3}$, B$_{3}$, C$_{3}$ and D$%
_{3}$ are dimensionless coefficients. The asymptotic factorization property
fixes the level 2 solution, i.e. the coefficients $b_{3}$, C$_{3}$ and D$%
_{3} $ to 
\begin{equation}
b_{3}=b=-\langle \Theta \rangle ^{2}\,,
\end{equation}
\begin{equation}
\text{ C}_{3}=-b\,,\text{ D}_{3}=-b\,.
\end{equation}
For $n=3$ the condition (\ref{boundstate}) fixes the level 1 part, i.e. the
coefficient B$_{3}$ to 
\begin{equation}
\text{B}_{3}=-b(1+2\cos 2\xi ).
\end{equation}
Now, let us consider the four-particle case. The most general level 2
solution to the form factor equations satisfying (\ref{factor}) and the
initial conditions 
\begin{equation}
F_{2}^{\mathcal{K}}=0  \label{K2=0}
\end{equation}
is 
\begin{equation}
\frac{b_{4}\sigma _{1}^{2}\sigma _{3}^{2}+\text{E}_{4}\sigma _{1}^{2}\sigma
_{2}\sigma _{4}+\text{D}_{4}\sigma _{2}\sigma _{3}^{2}+\text{C}_{4}\sigma
_{2}^{2}\sigma _{4}}{\sigma _{4}^{2}}F_{4}^{K_{4}}+\text{B}_{4}\,\frac{%
\sigma _{1}\sigma _{3}}{\sigma _{4}}F_{4}^{K_{4}}+\text{A}_{4}F_{4}^{K_{4}}%
\text{ .}  \label{cin-sol-K4}
\end{equation}
Here, $F_{4}^{K_{4}}$ is defined by 
\begin{equation*}
Q_{4}^{K_{4}}=(\sigma _{1}\sigma _{3}\sigma _{2}-\sigma _{3}^{2}-\sigma
_{1}^{2}\sigma _{4})\text{ }
\end{equation*}
and generates the one dimensional space of the level 0 solutions of the
equations (\ref{fn0})-(\ref{fn4}) which vanish at two particles. The
asymptotic factorization property fixes the level 2 part of (\ref{cin-sol-K4}%
), i.e. the coefficients $b_{4}$ to the same value of $b$, and C$_{4}$, D$%
_{4}$ and E$_{4}$ to 
\begin{equation}
\text{C}_{4}=-b\,,\text{ D}_{4}=0\,,\text{ E}_{4}=0\,\text{\ .}
\end{equation}
The general result (\ref{F(n+m)-vev-general}) for $n=m=2$ is rewritten as 
\begin{eqnarray}
&&\left. \lim_{\eta \rightarrow 0}F_{4}^{T\bar{T}}(\theta _{2}+\eta +i\pi
,\theta _{1}+\eta +i\pi ,\theta _{1},\theta _{2})=\,\lim_{\eta \rightarrow
0}\{[F_{2}^{T}(\theta _{1}+\eta +i\pi ,\theta _{1})F_{2}^{\bar{T}}(\theta
_{2}+\eta +i\pi ,\theta _{2})\right.  \notag \\
&&\left. -F_{2}^{\Theta }(\theta _{1}+\eta +i\pi ,\theta _{1})F_{2}^{\Theta
}(\theta _{2}+\eta +i\pi ,\theta _{2})-\langle \Theta \rangle F_{4}^{\Theta
}(\theta _{2}+\eta +i\pi ,\theta _{1}+\eta +i\pi ,\theta _{1},\theta
_{2})]\right.  \notag \\
&&\left. +[T\leftrightarrow \bar{T}]\}\,.\right.  \label{FF-TTbar(n)}
\end{eqnarray}
This expression implies for the four-particle kernel 
\begin{equation}
\lim_{\eta \rightarrow 0}F_{4}^{\mathcal{K}}(\theta _{2}+\eta +i\pi ,\theta
_{1}+\eta +i\pi ,\theta _{1},\theta _{2})=2[F_{2}^{\Theta }(i\pi
,0)]^{2}\,(\cosh 2(\theta _{1}-\theta _{2})-1)\,,
\end{equation}
where the r.h.s. comes from the identity 
\begin{equation}
F_{2}^{T}(\theta _{1}+i\pi ,\theta _{1})F_{2}^{\bar{T}}(\theta _{2}+i\pi
,\theta _{2})-F_{2}^{\Theta }(\theta _{1}+i\pi ,\theta _{1})F_{2}^{\Theta
}(\theta _{2}+i\pi ,\theta _{2})=[e^{2(\theta _{1}-\theta
_{2})}-1]\,[F_{2}^{\Theta }(i\pi ,0)]^{2}.
\end{equation}
The above condition fixes the level 0 part of (\ref{cin-sol-K4}), i.e. the
coefficient A$_{4}$ to 
\begin{equation}
\text{A}_{4}=4b\cos ^{2}\xi \text{ }.\,
\end{equation}
Finally, for $n=4$ the condition (\ref{boundstate}) fixes the level 1 part,
i.e. the coefficient B$_{4}$ to 
\begin{equation}
\text{B}_{4}=-2b\cos \xi (1+4\cos \xi +2\cos 2\xi ).
\end{equation}

This procedure has been implemented up to nine particles and the result is
given in the next appendix.

\resection{Appendix} \label{Ap-explicit-TTbar-up9} We list in this appendix
the functions $\tilde{Q}_{n}^{\mathcal{K}}$ which through (\ref{fn}), (\ref
{un}) and 
\begin{equation}
Q_{n}^{\mathcal{K}}=-\langle \Theta \rangle ^{2}\,\tilde{Q}_{n}^{\mathcal{K}}
\end{equation}
determine $F_{n}^{\mathcal{K}}$ in (\ref{fnttbar}). The $\tilde{Q}_{n}^{%
\mathcal{K}}$ have been determined explicitly up to $n=9$. The functions $%
\tilde{Q}_{8}^{\mathcal{K}}$ and $\tilde{Q}_{9}^{\mathcal{K}}$, however, are
too cumbersome and we do not reproduce them here. We simplify the notation
by dropping the superscript $(n)$ on the symmetric polynomials. 
\begin{equation}
\tilde{Q}_{3}^{\mathcal{K}}=\frac{1}{\sigma _{3}^{2}}(\sigma _{1}^{2}\sigma
_{2}^{2}-\sigma _{2}^{3}-\sigma _{1}^{3}\sigma _{3}-(1+2\cos [2\xi
])\,\sigma _{1}\sigma _{2}\sigma _{3})(\sigma _{1}\sigma _{2}-\sigma _{3})%
\text{ \ \ \ \ \ \ \ \ \ \ \ \ \ \ \ \ \ \ \ \ \ \ \ }
\end{equation}
\begin{eqnarray}
&&\left. \tilde{Q}_{4}^{\mathcal{K}}=\right. \frac{1}{\sigma _{4}^{2}}%
(\sigma _{1}^{2}\sigma _{3}^{2}-\sigma _{2}^{2}\sigma _{4}-2\cos [\xi
](1+4\cos [\xi ]+2\cos [2\xi ])\,\sigma _{1}\sigma _{3}\sigma _{4}+4\cos
[\xi ]^{2}\sigma _{4}^{2})\text{\ \ \ \ \ \ \ \ \ \ \ \ \ \ \ }  \notag \\
&&\text{ }(\sigma _{1}\sigma _{3}\sigma _{2}-\sigma _{3}^{2}-\sigma
_{1}^{2}\sigma _{4})\text{ \ \ \ \ \ \ \ \ \ \ \ \ \ \ \ \ \ \ \ \ \ \ \ \ \
\ \ \ \ \ \ \ \ \ \ \ \ \ \ \ \ \ \ \ \ \ \ \ \ \ \ \ \ \ \ \ \ \ \ \ \ \ \
\ \ \ \ \ \ \ \ \ \ \ \ \ \ \ \ \ \ \ \ \ \ \ \ \ \ \ \ \ \ \ \ \ }
\end{eqnarray}
\begin{eqnarray}
&&\left. \tilde{Q}_{5}^{\mathcal{K}}=\right. \frac{1}{\sigma _{5}^{2}}%
(-\sigma _{1}^{4}{{\sigma }_{3}}\sigma _{4}^{2}{{\sigma }_{5}}+\sigma
_{4}^{2}{{\sigma }_{5}}(\sigma _{2}^{2}{{\sigma }_{3}}-4{{\cos [\xi ]}^{2}}{{%
\sigma }_{3}}{{\sigma }_{4}}-{{\sigma }_{2}}{{\sigma }_{5}})+{{\sigma }_{1}}{%
{\sigma }_{4}}{{\sigma }_{5}}(-\sigma _{2}^{3}{{\sigma }_{4}}+{{\sigma }_{2}}%
(2(2+\cos [2\xi ])\sigma _{4}^{2}  \notag \\
&&-(7+8\cos [\xi ]+8\cos [2\xi ]+4\cos [3\xi ]+2\cos [4\xi ]){{\sigma }_{3}}{%
{\sigma }_{5}})+4\cos [\xi ]((1+4\cos [\xi ]+2\cos [2\xi ]  \notag \\
&&+\cos [3\xi ])\sigma _{3}^{2}{{\sigma }_{4}}+(1+2\cos [\xi ]+2\cos [2\xi
])\sigma _{5}^{2}))+\sigma _{1}^{2}(2(4+4\cos [\xi ]+5\cos [2\xi ]+2\cos
[3\xi ]  \notag \\
&&+\cos [4\xi ])\sigma _{2}^{2}{{\sigma }_{4}}\sigma _{5}^{2}-{{\sigma }_{5}}%
(\sigma _{3}^{3}{{\sigma }_{4}}+8{{\cos [\frac{\xi }{2}]}^{2}}(2+\cos [\xi
]+2\cos [2\xi ]+\cos [3\xi ])\sigma _{4}^{2}{{\sigma }_{5}}+{{\sigma }_{3}}%
\sigma _{5}^{2})  \notag \\
&&-{{\sigma }_{2}}(\sigma _{4}^{4}+2(5+4\cos [\xi ]+6\cos [2\xi ]+2\cos
[3\xi ]+\cos [4\xi ]){{\sigma }_{3}}\sigma _{4}^{2}{{\sigma }_{5}}-\sigma
_{3}^{2}\sigma _{5}^{2}))  \notag \\
&&+\sigma _{1}^{3}(2{{\sigma }_{4}}{{\sigma }_{5}}((3+4\cos [\xi ]+4\cos
[2\xi ]+2\cos [3\xi ]+\cos [4\xi ])\sigma _{4}^{2}+(2+\cos [2\xi ]){{\sigma }%
_{3}}{{\sigma }_{5}})  \notag \\
&&+{{\sigma }_{2}}({{\sigma }_{3}}\sigma _{4}^{3}-4{{\cos [\xi ]}^{2}}\sigma
_{5}^{3})))
\end{eqnarray}
\begin{eqnarray}
&&\left. \tilde{Q}_{6}^{\mathcal{K}}=\right. \frac{1}{\sigma _{6}^{2}}%
(\sigma _{1}^{3}{{\sigma }_{2}}{{\sigma }_{3}}{{\sigma }_{4}}\sigma
_{5}^{3}-\sigma _{1}^{2}{{\sigma }_{2}}{{\sigma }_{3}}\sigma _{5}^{4}-4{{%
\cos [\xi ]}^{2}}\sigma _{1}^{3}{{\sigma }_{4}}\sigma _{5}^{4}+4{{\cos [\xi ]%
}^{2}}\sigma _{1}^{2}\sigma _{5}^{5}-\sigma _{1}^{2}{{\sigma }_{2}}\sigma
_{4}^{3}{{\sigma }_{5}}{{\sigma }_{6}}-\sigma _{1}^{2}\sigma _{3}^{3}\sigma
_{5}^{2}{{\sigma }_{6}}  \notag \\
&&-{{\sigma }_{1}}\sigma _{2}^{3}{{\sigma }_{4}}\sigma _{5}^{2}{{\sigma }_{6}%
}-\sigma _{1}^{4}{{\sigma }_{3}}{{\sigma }_{4}}\sigma _{5}^{2}{{\sigma }_{6}}%
-2\cos [\xi ](7+16\cos [\xi ]+12\cos [2\xi ]+4\cos [3\xi ]  \notag \\
&&+2\cos [4\xi ])\sigma _{1}^{2}{{\sigma }_{2}}{{\sigma }_{3}}{{\sigma }_{4}}%
\sigma _{5}^{2}{{\sigma }_{6}}+\sigma _{2}^{2}{{\sigma }_{3}}{{\sigma }_{4}}%
\sigma _{5}^{2}{{\sigma }_{6}}+2(6+11\cos [\xi ]+8\cos [2\xi ]+6\cos [3\xi
]+2\cos [4\xi ]  \notag \\
&&+\cos [5\xi ]){{\sigma }_{1}}\sigma _{3}^{2}{{\sigma }_{4}}\sigma _{5}^{2}{%
{\sigma }_{6}}+2(6+11\cos [\xi ]+8\cos [2\xi ]+6\cos [3\xi ]+2\cos [4\xi
]+\cos [5\xi ])\sigma _{1}^{3}\sigma _{4}^{2}\sigma _{5}^{2}{{\sigma }_{6}} 
\notag \\
&&+2(2+\cos [2\xi ]){{\sigma }_{1}}{{\sigma }_{2}}\sigma _{4}^{2}\sigma
_{5}^{2}{{\sigma }_{6}}-4{{\cos [\xi ]}^{2}}{{\sigma }_{3}}\sigma
_{4}^{2}\sigma _{5}^{2}{{\sigma }_{6}}-4{{\cos [\xi ]}^{2}}\sigma _{1}^{4}{{%
\sigma }_{2}}\sigma _{5}^{3}{{\sigma }_{6}}+2(6+11\cos [\xi ]+8\cos [2\xi ] 
\notag \\
&&+6\cos [3\xi ]+2\cos [4\xi ]+\cos [5\xi ])\sigma _{1}^{2}\sigma
_{2}^{2}\sigma _{5}^{3}{{\sigma }_{6}}+(5+4\cos [2\xi ])\sigma _{1}^{3}{{%
\sigma }_{3}}\sigma _{5}^{3}{{\sigma }_{6}}-2{{(1+2\cos [\xi ])}^{3}}(1-\cos
[\xi ]  \notag \\
&&+\cos [2\xi ]){{\sigma }_{1}}{{\sigma }_{2}}{{\sigma }_{3}}\sigma _{5}^{3}{%
{\sigma }_{6}}-2(6+15\cos [\xi ]+8\cos [2\xi ]+8\cos [3\xi ]+2\cos [4\xi
]+\cos [5\xi ])\sigma _{1}^{2}{{\sigma }_{4}}\sigma _{5}^{3}{{\sigma }_{6}} 
\notag \\
&&-{{\sigma }_{2}}{{\sigma }_{4}}\sigma _{5}^{3}{{\sigma }_{6}}+4(2\cos [\xi
]+\cos [3\xi ]){{\sigma }_{1}}\sigma _{5}^{4}{{\sigma }_{6}}+\sigma _{1}^{2}{%
{\sigma }_{2}}{{\sigma }_{3}}\sigma _{4}^{2}\sigma _{6}^{2}+2(6+11\cos [\xi
]+8\cos [2\xi ]+6\cos [3\xi ]  \notag \\
&&+2\cos [4\xi ]+\cos [5\xi ])\sigma _{1}^{2}{{\sigma }_{2}}\sigma _{3}^{2}{{%
\sigma }_{5}}\sigma _{6}^{2}-2{{(1+2\cos [\xi ])}^{3}}(1-\cos [\xi ]+\cos
[2\xi ]){{\sigma }_{1}}\sigma _{3}^{3}{{\sigma }_{5}}\sigma _{6}^{2}+2(2 
\notag \\
&&+\cos [2\xi ])\sigma _{1}^{2}\sigma _{2}^{2}{{\sigma }_{4}}{{\sigma }_{5}}%
\sigma _{6}^{2}-2{{(1+2\cos [\xi ])}^{3}}(1-\cos [\xi ]+\cos [2\xi ])\sigma
_{1}^{3}{{\sigma }_{3}}{{\sigma }_{4}}{{\sigma }_{5}}\sigma _{6}^{2}-2{{%
\sigma }_{1}}{{\sigma }_{2}}{{\sigma }_{3}}{{\sigma }_{4}}{{\sigma }_{5}}%
\sigma _{6}^{2}  \notag \\
&&-(3+4\cos [2\xi ])\sigma _{1}^{2}\sigma _{4}^{2}{{\sigma }_{5}}\sigma
_{6}^{2}+4{{\cos [\xi ]}^{2}}\sigma _{1}^{5}\sigma _{5}^{2}\sigma
_{6}^{2}-2(6+15\cos [\xi ]+8\cos [2\xi ]+8\cos [3\xi ]+2\cos [4\xi ]  \notag
\\
&&+\cos [5\xi ])\sigma _{1}^{3}{{\sigma }_{2}}\sigma _{5}^{2}\sigma
_{6}^{2}-(3+4\cos [2\xi ]){{\sigma }_{1}}\sigma _{2}^{2}\sigma
_{5}^{2}\sigma _{6}^{2}+2(4+17\cos [\xi ]+6\cos [2\xi ]+9\cos [3\xi ]+2\cos
[4\xi ]  \notag \\
&&+\cos [5\xi ])\sigma _{1}^{2}{{\sigma }_{3}}\sigma _{5}^{2}\sigma
_{6}^{2}+(1+2\cos [2\xi ]){{\sigma }_{2}}{{\sigma }_{3}}\sigma
_{5}^{2}\sigma _{6}^{2}-2(1+\cos [2\xi ]+\cos [4\xi ]){{\sigma }_{1}}{{%
\sigma }_{4}}\sigma _{5}^{2}\sigma _{6}^{2}+{(1}  \notag \\
&&{{+2\cos [2\xi ])}^{2}}\sigma _{5}^{3}\sigma _{6}^{2}-4{{\cos [\xi ]}^{2}}%
\sigma _{1}^{2}\sigma _{2}^{2}{{\sigma }_{3}}\sigma _{6}^{3}-\sigma _{1}^{3}{%
{\sigma }_{2}}{{\sigma }_{4}}\sigma _{6}^{3}+(1+2\cos [2\xi ])\sigma _{1}^{2}%
{{\sigma }_{3}}{{\sigma }_{4}}\sigma _{6}^{3}+4(2\cos [\xi ]  \notag \\
&&+\cos [3\xi ])\sigma _{1}^{4}{{\sigma }_{5}}\sigma _{6}^{3}-2(1+\cos [2\xi
]+\cos [4\xi ])\sigma _{1}^{2}{{\sigma }_{2}}{{\sigma }_{5}}\sigma _{6}^{3}+2%
{{(1+2\cos [2\xi ])}^{2}}{{\sigma }_{1}}{{\sigma }_{3}}{{\sigma }_{5}}\sigma
_{6}^{3}  \notag \\
&&+{{(1+2\cos [2\xi ])}^{2}}\sigma _{1}^{3}\sigma _{6}^{4})
\end{eqnarray}
\begin{eqnarray}
&&\left. \tilde{Q}_{7}^{\mathcal{K}}=\right. \frac{1}{\sigma _{7}^{2}}%
(\sigma _{1}^{5}\sigma _{6}^{2}{{\sigma }_{7}}(4{{\cos [\xi ]}^{2}}{{\sigma }%
_{5}}{{\sigma }_{6}}-(1+2\cos [2\xi ]){{\sigma }_{4}}{{\sigma }_{7}})-\sigma
_{1}^{4}{{\sigma }_{6}}(-{{\sigma }_{7}}(-{{\sigma }_{3}}{{\sigma }_{4}}{{%
\sigma }_{5}}{{\sigma }_{6}}+(3+8\cos [\xi ]  \notag \\
&&+4\cos [2\xi ]+4\cos [3\xi ]+2\cos [4\xi ])\sigma _{6}^{3}-2(6+7\cos [2\xi
]+2\cos [4\xi ]){{\sigma }_{5}}{{\sigma }_{6}}{{\sigma }_{7}}  \notag \\
&&+8{{\cos [\xi ]}^{2}}(1+2\cos [2\xi ]){{\sigma }_{4}}\sigma _{7}^{2})+{{%
\sigma }_{2}}{{\sigma }_{6}}(4{{\cos [\xi ]}^{2}}{{\sigma }_{5}}\sigma
_{6}^{2}-{{\sigma }_{7}}((1+2\cos [2\xi ]){{\sigma }_{4}}{{\sigma }_{6}}+4{{%
\cos [\xi ]}^{2}}{{\sigma }_{3}}{{\sigma }_{7}})))  \notag \\
&&\text{-}\sigma _{6}^{2}{{\sigma }_{7}}(4{{\cos [\xi ]}^{2}}\sigma _{2}^{3}{%
{\sigma }_{3}}{{\sigma }_{7}}\text{-}\sigma _{2}^{2}{{\sigma }_{5}}({{\sigma 
}_{3}}{{\sigma }_{4}}+4{{\cos [\xi ]}^{2}}{{\sigma }_{7}})\text{-}{{\sigma }%
_{5}}({{(1+2\cos [2\xi ])}^{2}}{{\sigma }_{5}}{{\sigma }_{6}}\text{-}8{{\cos
[\xi ]}^{2}}\cos [2\xi ]{{\sigma }_{4}}{{\sigma }_{7}})  \notag \\
&&+4{{\cos [\xi ]}^{2}}{{\sigma }_{3}}(\sigma _{4}^{2}{{\sigma }_{5}}+4{{%
\cos [\xi ]}^{2}}(1+2\cos [2\xi ]){{\sigma }_{6}}{{\sigma }_{7}})+{{\sigma }%
_{2}}(-(1+2\cos [2\xi ]){{\sigma }_{3}}{{\sigma }_{5}}{{\sigma }_{6}}+4{{%
\cos [\xi ]}^{2}}\sigma _{7}^{2}  \notag \\
&&+{{\sigma }_{4}}(\sigma _{5}^{2}-16{{\cos [\xi ]}^{4}}{{\sigma }_{3}}{{%
\sigma }_{7}})))+{{\sigma }_{1}}{{\sigma }_{6}}{{\sigma }_{7}}(-\sigma
_{2}^{3}{{\sigma }_{4}}{{\sigma }_{5}}{{\sigma }_{6}}-4{{\cos [\xi ]}^{2}}%
(-1+2\cos [\xi ]){{(1+2\cos [\xi ])}^{3}}\sigma _{3}^{3}\sigma _{6}^{2} 
\notag \\
&&+4{{\cos [\xi ]}^{2}}\sigma _{2}^{4}{{\sigma }_{6}}{{\sigma }_{7}}-4\cos
[\xi ]\sigma _{3}^{2}{{\sigma }_{4}}(-(5+13\cos [\xi ]+8\cos [2\xi ]+6\cos
[3\xi ]+2\cos [4\xi ]+\cos [5\xi ]){{\sigma }_{5}}{{\sigma }_{6}}  \notag \\
&&+{{(1+2\cos [\xi ])}^{3}}(1-\cos [\xi ]+\cos [2\xi ]){{\sigma }_{4}}{{%
\sigma }_{7}})+(1+2\cos [2\xi ]){{\sigma }_{3}}(8{{\cos [\xi ]}^{2}}\sigma
_{6}^{3}+(11+20\cos [2\xi ]  \notag \\
&&+8\cos [3\xi ]+10\cos [4\xi ]+4\cos [5\xi ]+2\cos [6\xi ]){{\sigma }_{5}}{{%
\sigma }_{6}}{{\sigma }_{7}}-8{{\cos [\xi ]}^{2}}(1+3\cos [2\xi ]+2\cos
[3\xi ]  \notag \\
&&+\cos [4\xi ]){{\sigma }_{4}}\sigma _{7}^{2})-\sigma _{2}^{2}((3+4\cos
[2\xi ]){{\sigma }_{5}}\sigma _{6}^{2}+{{\sigma }_{7}}((9+10\cos [2\xi
]+2\cos [4\xi ]){{\sigma }_{4}}{{\sigma }_{6}}-8{{\cos [\xi ]}^{2}}{{\sigma }%
_{3}}{{\sigma }_{7}}))  \notag \\
&&+{{\sigma }_{2}}(2(2+\cos [2\xi ])\sigma _{4}^{2}{{\sigma }_{5}}{{\sigma }%
_{6}}-4{{\cos [\xi ]}^{2}}(-1+2\cos [\xi ]){{(1+2\cos [\xi ])}^{3}}{{\sigma }%
_{3}}\sigma _{5}^{2}{{\sigma }_{6}}+(11+20\cos [\xi ]  \notag \\
&&+20\cos [2\xi ]+12\cos [3\xi ]+10\cos [4\xi ]+4\cos [5\xi ]+2\cos [6\xi ]){%
{\sigma }_{3}}{{\sigma }_{4}}{{\sigma }_{5}}{{\sigma }_{7}}+8{{\cos [\xi ]}%
^{2}}(1+4\cos [\xi ]  \notag \\
&&+3\cos [2\xi ]+2\cos [3\xi ]+\cos [4\xi ])\sigma _{3}^{2}{{\sigma }_{6}}{{%
\sigma }_{7}}-{{\sigma }_{7}}(-(19+26\cos [2\xi ]+10\cos [4\xi ]+2\cos [6\xi
])\sigma _{6}^{2}  \notag \\
&&+4(3+5\cos [\xi ]+4\cos [2\xi ]+3\cos [3\xi ]+\cos [4\xi ]+\cos [5\xi ]){{%
\sigma }_{5}}{{\sigma }_{7}}))+2(2(2\cos [\xi ]+\cos [3\xi ])\sigma _{5}^{3}{%
{\sigma }_{6}}  \notag \\
&&-(3+3\cos [2\xi ]+\cos [4\xi ]){{\sigma }_{4}}{{\sigma }_{5}}\sigma
_{6}^{2}+4{{\cos [\xi ]}^{2}}{{\sigma }_{4}}\sigma _{5}^{2}{{\sigma }_{7}}+2{%
{\cos [\xi ]}^{2}}(1+2\cos [2\xi ]){{\sigma }_{7}}(\sigma _{4}^{2}{{\sigma }%
_{6}}+2(1  \notag \\
&&+2\cos [\xi ]+2\cos [2\xi ]+2\cos [3\xi ])\sigma _{7}^{2})))+\sigma
_{1}^{3}(2(13+18\cos [\xi ]+19\cos [2\xi ]+10\cos [3\xi ]+7\cos [4\xi ] 
\notag \\
&&+2\cos [5\xi ]+\cos [6\xi ])\sigma _{4}^{2}{{\sigma }_{5}}\sigma _{6}^{2}{{%
\sigma }_{7}}-4{{\cos [\xi ]}^{2}}(-1+2\cos [\xi ]){{(1+2\cos [\xi ])}^{3}}%
\sigma _{4}^{3}{{\sigma }_{6}}\sigma _{7}^{2}  \notag \\
&&+\sigma _{2}^{2}(-4{{\cos [\xi ]}^{2}}{{\sigma }_{3}}\sigma _{6}^{3}{{%
\sigma }_{7}}+{{(1+2\cos [2\xi ])}^{2}}\sigma _{7}^{4})+{{\sigma }_{6}}%
(-(1+2\cos [2\xi ])\sigma _{3}^{2}{{\sigma }_{6}}\sigma _{7}^{2}+{{\sigma }%
_{3}}{{\sigma }_{5}}{{\sigma }_{6}}((1  \notag \\
&&+2\cos [2\xi ])\sigma _{6}^{2}+2(2+\cos [2\xi ]){{\sigma }_{5}}{{\sigma }%
_{7}})+\sigma _{7}^{2}(2(13+14\cos [\xi ]+22\cos [2\xi ]+12\cos [3\xi ] 
\notag \\
&&+13\cos [4\xi ]+8\cos [5\xi ]+5\cos [6\xi ]+2\cos [7\xi ]+\cos [8\xi
])\sigma _{6}^{2}+(19+26\cos [2\xi ]+10\cos [4\xi ]  \notag \\
&&+2\cos [6\xi ]){{\sigma }_{5}}{{\sigma }_{7}}))-4{{\cos [\xi ]}^{2}}{{%
\sigma }_{4}}(\sigma _{5}^{2}\sigma _{6}^{3}+(1+2\cos [2\xi ]){{\sigma }_{7}}%
({{(1+2\cos [\xi ])}^{2}}{{\sigma }_{3}}\sigma _{6}^{3}+4{{\cos [\xi ]}^{2}}%
\sigma _{7}^{3}))  \notag \\
&&+{{\sigma }_{2}}(4{{\cos [\xi ]}^{2}}\sigma _{6}^{5}-2(7+22\cos [\xi
]+11\cos [2\xi ]+12\cos [3\xi ]+5\cos [4\xi ]+2\cos [5\xi ]  \notag \\
&&+\cos [6\xi ]){{\sigma }_{5}}\sigma _{6}^{3}{{\sigma }_{7}}-(3+4\cos [2\xi
])\sigma _{5}^{2}{{\sigma }_{6}}\sigma _{7}^{2}+2(7+22\cos [\xi ]+11\cos
[2\xi ]+12\cos [3\xi ]+5\cos [4\xi ]  \notag \\
&&+2\cos [5\xi ]+\cos [6\xi ]){{\sigma }_{4}}\sigma _{6}^{2}\sigma
_{7}^{2}+(1+2\cos [2\xi ]){{\sigma }_{4}}{{\sigma }_{5}}\sigma _{7}^{3}+{{%
\sigma }_{3}}{{\sigma }_{6}}({{\sigma }_{4}}{{\sigma }_{5}}\sigma
_{6}^{2}-2(3+3\cos [2\xi ]  \notag \\
&&+\cos [4\xi ])\sigma _{7}^{3})))+\sigma _{1}^{2}(-{{\sigma }_{3}}\sigma
_{6}^{5}-2\cos [2\xi ]{{\sigma }_{3}}\sigma _{6}^{5}+2\sigma _{5}^{4}{{%
\sigma }_{6}}{{\sigma }_{7}}+2\cos [2\xi ]\sigma _{5}^{4}{{\sigma }_{6}}{{%
\sigma }_{7}}-\sigma _{3}^{3}{{\sigma }_{5}}\sigma _{6}^{2}{{\sigma }_{7}} 
\notag \\
&&+14{{\sigma }_{3}}{{\sigma }_{5}}\sigma _{6}^{3}{{\sigma }_{7}}+44\cos
[\xi ]{{\sigma }_{3}}{{\sigma }_{5}}\sigma _{6}^{3}{{\sigma }_{7}}+22\cos
[2\xi ]{{\sigma }_{3}}{{\sigma }_{5}}\sigma _{6}^{3}{{\sigma }_{7}}+24\cos
[3\xi ]{{\sigma }_{3}}{{\sigma }_{5}}\sigma _{6}^{3}{{\sigma }_{7}}  \notag
\\
&&+10\cos [4\xi ]{{\sigma }_{3}}{{\sigma }_{5}}\sigma _{6}^{3}{{\sigma }_{7}}%
+4\cos [5\xi ]{{\sigma }_{3}}{{\sigma }_{5}}\sigma _{6}^{3}{{\sigma }_{7}}%
+2\cos [6\xi ]{{\sigma }_{3}}{{\sigma }_{5}}\sigma _{6}^{3}{{\sigma }_{7}}-9{%
{\sigma }_{3}}\sigma _{5}^{2}{{\sigma }_{6}}\sigma _{7}^{2}  \notag \\
&&-10\cos [2\xi ]{{\sigma }_{3}}\sigma _{5}^{2}{{\sigma }_{6}}\sigma
_{7}^{2}-2\cos [4\xi ]{{\sigma }_{3}}\sigma _{5}^{2}{{\sigma }_{6}}\sigma
_{7}^{2}+4(2\cos [\xi ]+\cos [3\xi ])\sigma _{2}^{3}{{\sigma }_{6}}\sigma
_{7}^{3}+4\sigma _{3}^{2}{{\sigma }_{6}}\sigma _{7}^{3}  \notag \\
&&+6\cos [2\xi ]\sigma _{3}^{2}{{\sigma }_{6}}\sigma _{7}^{3}+2\cos [4\xi
]\sigma _{3}^{2}{{\sigma }_{6}}\sigma _{7}^{3}-49\sigma _{6}^{2}\sigma
_{7}^{3}-80\cos [\xi ]\sigma _{6}^{2}\sigma _{7}^{3}-80\cos [2\xi ]\sigma
_{6}^{2}\sigma _{7}^{3}  \notag \\
&&-60\cos [3\xi ]\sigma _{6}^{2}\sigma _{7}^{3}-44\cos [4\xi ]\sigma
_{6}^{2}\sigma _{7}^{3}-32\cos [5\xi ]\sigma _{6}^{2}\sigma _{7}^{3}-14\cos
[6\xi ]\sigma _{6}^{2}\sigma _{7}^{3}-8\cos [7\xi ]\sigma _{6}^{2}\sigma
_{7}^{3}  \notag \\
&&-2\cos [8\xi ]\sigma _{6}^{2}\sigma _{7}^{3}-2{{\sigma }_{5}}\sigma
_{7}^{4}-2\cos [2\xi ]{{\sigma }_{5}}\sigma _{7}^{4}-\sigma _{4}^{2}{{\sigma 
}_{6}}{{\sigma }_{7}}((1+2\cos [2\xi ])\sigma _{6}^{2}-8{{\cos [\xi ]}^{2}}%
(1+4\cos [\xi ]  \notag \\
&&+3\cos [2\xi ]+2\cos [3\xi ]+\cos [4\xi ]){{\sigma }_{5}}{{\sigma }_{7}}%
)+\sigma _{2}^{2}{{\sigma }_{7}}(2(13+18\cos [\xi ]+19\cos [2\xi ]+10\cos
[3\xi ]+7\cos [4\xi ]  \notag \\
&&+2\cos [5\xi ]+\cos [6\xi ])\sigma _{5}^{2}\sigma _{6}^{2}-{{\sigma }_{5}}{%
{\sigma }_{7}}(4{{\cos [\xi ]}^{2}}(-1+2\cos [\xi ]){{(1+2\cos [\xi ])}^{3}}{%
{\sigma }_{4}}{{\sigma }_{6}}+{{\sigma }_{3}}{{\sigma }_{7}})+2\sigma
_{6}^{2}((2  \notag \\
&&+\cos [2\xi ]){{\sigma }_{4}}{{\sigma }_{6}}-(10+22\cos [\xi ]+15\cos
[2\xi ]+12\cos [3\xi ]+6\cos [4\xi ]+2\cos [5\xi ]+\cos [6\xi ]){{\sigma }%
_{3}}{{\sigma }_{7}}))  \notag \\
&&-4\cos [\xi ]{{\sigma }_{4}}((6+10\cos [\xi ]+10\cos [2\xi ]+5\cos [3\xi
]+2\cos [4\xi ]+\cos [5\xi ])\sigma _{5}^{2}\sigma _{6}^{2}{{\sigma }_{7}}%
+\cos [\xi ]\sigma _{5}^{3}\sigma _{7}^{2}  \notag \\
&&+(1+2\cos [2\xi ]){{\sigma }_{3}}\sigma _{6}^{2}\sigma _{7}^{2}-\cos [\xi ]%
{{\sigma }_{5}}(\sigma _{6}^{4}+4{{\cos [\xi ]}^{2}}{{\sigma }_{3}}\sigma
_{7}^{3}))+{{\sigma }_{2}}(2\sigma _{3}^{2}{{\sigma }_{7}}((13+18\cos [\xi
]+19\cos [2\xi ]  \notag
\end{eqnarray}
\newpage

\begin{eqnarray}
&&+10\cos [3\xi ]+7\cos [4\xi ]+2\cos [5\xi ]+\cos [6\xi ])\sigma
_{6}^{3}+(2+\cos [2\xi ]){{\sigma }_{5}}{{\sigma }_{6}}{{\sigma }_{7}}-2{{%
\cos [\xi ]}^{2}}{{\sigma }_{4}}\sigma _{7}^{2})+{{\sigma }_{7}}(-\sigma
_{4}^{3}\sigma _{6}^{2}  \notag \\
&&+(31+8\cos [\xi ]+52\cos [2\xi ]+12\cos [3\xi ]+32\cos [4\xi ]+12\cos
[5\xi ]+12\cos [6\xi ]+4\cos [7\xi ]  \notag \\
&&+2\cos [8\xi ]){{\sigma }_{4}}{{\sigma }_{6}}\sigma _{7}^{2}-2((6+7\cos
[2\xi ]+2\cos [4\xi ])\sigma _{6}^{4}+8{{\cos [\xi ]}^{3}}(2\cos [\xi
]-2\cos [2\xi ]  \notag \\
&&+3\cos [3\xi ]+2\cos [4\xi ]+\cos [5\xi ]){{\sigma }_{5}}\sigma _{6}^{2}{{%
\sigma }_{7}}-2{{\cos [\xi ]}^{2}}\sigma _{5}^{2}\sigma _{7}^{2}))+{{\sigma }%
_{3}}(2(13+18\cos [\xi ]  \notag \\
&&+19\cos [2\xi ]+10\cos [3\xi ]+7\cos [4\xi ]+2\cos [5\xi ]+\cos [6\xi
])\sigma _{4}^{2}{{\sigma }_{6}}\sigma _{7}^{2}-{{\sigma }_{4}}(\sigma
_{6}^{4}+2(18+22\cos [\xi ]  \notag \\
&&+25\cos [2\xi ]+12\cos [3\xi ]+8\cos [4\xi ]+2\cos [5\xi ]+\cos [6\xi ]){{%
\sigma }_{5}}\sigma _{6}^{2}{{\sigma }_{7}}-\sigma _{5}^{2}\sigma _{7}^{2})-{%
{\sigma }_{7}}(\sigma _{5}^{3}{{\sigma }_{6}}  \notag \\
&&+8{{\cos [\xi ]}^{2}}\cos [2\xi ]\sigma _{7}^{3})))))
\end{eqnarray}

\resection{Appendix} \label{Ap-M(2,2N+3)-boundstate}In this appendix we
explicitly derive the form factor equations which characterize the operator
content of the $\Phi _{1,3}$-perturbed minimal model $\mathcal{M}_{2,2N+3}$.
The particles in such models obey the following bootstrap fusion algebra 
\begin{equation}
B_{a}\times B_{b}\rightarrow B_{\min (a+b,2N+1-a-b)}\text{\ and \ }%
B_{a}\times B_{b}\rightarrow B_{\left| a-b\right| }\,,\hspace{1cm}%
a,b=1,\ldots,N  \label{fusion-particles}
\end{equation}
which correspond to the bound state poles located at 
\begin{align}
& \left. \theta _{ab}^{\min (a+b,2N+1-a-b)}=i(a+b)\frac{\xi _{N}}{2}\right. 
\text{ } \\
& \left. \theta _{ab}^{|a-b|}=i(\pi -|a-b|\frac{\xi _{N}}{2})\text{ , \ }%
a\neq b\text{ .}\right.
\end{align}
The pattern (\ref{fusion-particles}) makes clear the possibility to describe
all the particles and their fusions in terms of the fusion processes
involving only the lightest particle $B_{1}$. In this way all the bound
state constraints can be handled inside the parametrization (\ref{fn}).

In particular, we can describe the particle $B_{\min (n,2N+1-n)}$ by the
fusion of $n$ $B_{1}$ particles set in pairs on the bound state pole of the $%
B_{1}B_{1}$ scattering channel. For the form factors of the generic local
operator $\Phi $ this implies 
\begin{eqnarray}
&&\left. \langle 0|\Phi (0)|B_{\min (n,2N+1-n)}(\tilde{\theta}%
_{n})B_{1}(\theta _{1}^{\prime })\ldots B_{1}(\theta _{m}^{\prime })\rangle
\,=\right.  \notag \\
&&\left. \frac{(-i)^{n-1}}{\Upsilon _{n}}\lim_{\epsilon _{1}\rightarrow
0}...\lim_{\epsilon _{n-1}\rightarrow 0}\epsilon _{1}\cdot \cdot \epsilon
_{n-1}F_{m+n}^{\Phi }(\theta _{1}+\epsilon _{1},..,\theta _{n-1}+\epsilon
_{n-1},\theta _{n},\theta _{1}^{\prime },..,\theta _{m}^{\prime })\right. 
\text{ }  \label{fusion-particles-ff}
\end{eqnarray}
for $1\leq n\leq 2N$, $\theta _{h}=\theta _{1}+i(h-1)\xi _{N}$, $h=2,\ldots
,n$, and $\tilde{\theta}_{n}=\theta _{1}+i(n-1)\xi _{N}/2$. The result does
not depend on the way in which the limit is done and $\Upsilon _{n}$ can be
written as 
\begin{equation}
\Upsilon _{n}=\left\{ 
\begin{array}{c}
\prod_{k=1}^{n}\Gamma _{1k}^{k+1}\text{ \ \ \ \ \ \ \ \ \ \ \ \ \ \ \ \ \ \
for }1\leq n\leq N\ , \\ 
\prod_{k=1}^{N}\Gamma _{1k}^{k+1}\prod_{h=2N+2-n}^{N}\Gamma _{1h}^{h-1}\text{
\ \ for }N<n\leq 2N,
\end{array}
\right. \text{ ,}
\end{equation}
where $\Gamma _{ab}^{c}$ is the three-particle coupling defined by 
\begin{equation}
\mbox{Res}_{\theta =\theta _{ab}^{c}}S_{ab}(\theta )=i\left( \Gamma
_{ab}^{c}\right) ^{2}\,.
\end{equation}
The use of formula (\ref{fusion-particles-ff}) allows to express the
conditions (\ref{B(2N+1-n)=B(n)})\ as 
\begin{align}
& \left. \lim_{\epsilon _{1}\rightarrow 0}...\lim_{\epsilon
_{2N-n}\rightarrow 0}\epsilon _{1}\cdot \cdot \epsilon
_{2N-n}F_{2N+1-n}^{\Phi }(\theta _{1}^{\prime }+\epsilon _{1},..,\theta
_{2N-n}^{\prime }+\epsilon _{2N-n},\theta _{2N+1-n}^{\prime })=\right. 
\notag \\
& \left. (-1)^{N-n}\frac{\Upsilon _{n}}{\Upsilon _{2N+1-n}}\lim_{\epsilon
_{1}\rightarrow 0}...\lim_{\epsilon _{n-1}\rightarrow 0}\epsilon _{1}\cdot
\cdot \epsilon _{n-1}F_{n}^{\Phi }(\theta _{1}+\epsilon _{1},..,\theta
_{n-1}+\epsilon _{n-1},\theta _{n})\right. ,  \label{FFB(2p+1-m)=FFB(m)}
\end{align}
with $\theta _{h}$ defined as above, $1\leq n\leq N,$ $\theta _{h}^{\prime
}=\theta _{1}^{\prime }+i(h-1)\xi _{N}$, $h=2,\ldots ,2N+1-n$, and $\theta
_{1}^{\prime }=\theta _{1}-i(2(N-n)+1)\xi _{N}/2$. Moreover, for each $m+2N$ 
$B_{1}$ particles with $m\geq 0$, we can consider the case $n=2N$ in (\ref
{fusion-particles-ff}) which implies the equation \cite{Koubek} 
\begin{align}
& \left. \lim_{\epsilon _{1}\rightarrow 0,..,\epsilon _{2N-1}\rightarrow
0}\epsilon _{1}\cdot \cdot \epsilon _{2N-1}F_{m+2N}^{\Phi }(\theta +\frac{%
2N-1}{2}\xi _{N}+\epsilon _{1},..,\theta -\frac{2N-1}{2}\xi _{N},\theta
_{1},..,\theta _{m})=\right.  \notag \\
& \left. \text{ \ \ \ \ \ \ \ \ \ \ \ \ \ \ \ \ \ \ \ \ \ \ \ \ \ \ \ \ \ \
\ \ \ \ \ \ \ \ \ \ \ \ \ \ \ \ \ \ \ \ \ \ \ \ \ \ \ \ \ \ \ \ \ \ \ \ \ }%
i^{(2N-1)}\Upsilon _{2N}F_{m+1}^{\Phi }(\theta ,\theta _{1},..,\theta _{m})%
\text{ .}\right.  \label{bound-state-eq-M(2,2p+3)}
\end{align}
In terms of the parametrization (\ref{fn}) the above equations become 
\begin{equation}
Q_{2N+1-n}^{\Phi }(\theta _{1}^{\prime },..,\theta _{2N+1-n}^{\prime
})=W_{n}^{N}(\theta _{1})Q_{n}^{\Phi }(\theta _{1},..,\theta _{n})\text{ },
\label{B(2p+1-m)=B(m)}
\end{equation}
\begin{equation}
W_{n}^{N}(\theta _{1})=\frac{Q_{2N+1-n}^{(1)}(\theta _{1}^{\prime
},..,\theta _{2N+1-n}^{\prime })}{Q_{n}^{(1)}(\theta _{1},\ldots ,\theta_{n})%
}
\end{equation}
and 
\begin{align}
& \left. Q_{m+2N}^{\Phi }(\theta +\frac{2N-1}{2}\xi _{N},\dots ,\theta +%
\frac{1}{2}\xi _{N},\theta -\frac{1}{2}\xi _{N},\dots ,\theta -\frac{2N-1}{2}%
\xi _{N},\theta _{1},\dots ,\theta _{m})=\right.  \notag \\
& \text{\ \ \ \ \ \ \ \ \ \ \ \ \ \ \ \ \ \ \ \ \ \ \ \ \ \ \ \ \ \ \ \ \ \
\ \ \ \ \ \ \ \ \ \ \ \ \ \ \ \ \ \ \ \ \ }V_{m}^{N}(x,x_{1},\dots
,x_{m})Q_{m+1}^{\Phi }(\theta ,\theta _{1},\dots ,\theta _{m})\text{ },
\label{B1-bound-state}
\end{align}
with 
\begin{equation}
V_{m}^{N}(x,x_{1},\dots ,x_{m})=(-1)^{\frac{N(N+1)}{2}+1}%
\prod_{k=2}^{N-1}[k]^{2}x^{N(2N-1)}\prod_{i=1}^{m}(x+x_{i})\prod_{k=2}^{N}%
\prod_{i=1}^{m}(x-x_{i}\omega ^{k})(x-x_{i}\omega ^{-k})\text{ },
\end{equation}
$x_{i}=e^{i\theta _{i}}$ and $\omega =e^{i\xi _{N}}$.

These bound state equations complete the equations (\ref{q1})-(\ref{q4}) and
characterize the operator content of this model. Let us describe the case of
the primary fields. For $\xi _{N}=2\pi /(2N+1)$ a periodicity arises and the
set of exponential operators $e^{ik\beta \varphi }$ of the sine-Gordon model
reduces to a finite set of $2N$ independent ones. Indeed, the form factors
of $e^{ik\beta \varphi }$ satisfy 
\begin{equation}
F_{n}^{k+(2N+1)}\left( \theta _{1},..,\theta _{n}\right) =F_{n}^{k}\left(
\theta _{1},..,\theta _{n}\right) \text{ ,}
\end{equation}
\begin{equation}
F_{n}^{(2N+1)-k}\left( \theta _{1},..,\theta _{n}\right) =\left( -1\right)
^{n+1}F_{n}^{k}\left( \theta _{1},..,\theta _{n}\right) \text{ ,}
\label{F(2n+1-a)=+-F(a)}
\end{equation}
implying that the independent among the operators $e^{ik\beta \varphi }$ can
be chosen for $k=1,..,2N$. Now, only those with $k=1,..,N$ have form factors
satisfying the bound state equations (\ref{B(2p+1-m)=B(m)}) and (\ref
{B1-bound-state}). Thus, the reduction of the operator content of the $\Phi
_{1,3}$-perturbed minimal model $\mathcal{M}_{2,2N+3}$ is implemented simply
requiring the bound state equations.

The form factors of the operator $T\bar{T}$ that we computed explicitly up
to nine particles satisfy the bound state equations (\ref{B(2p+1-m)=B(m)})
and (\ref{B1-bound-state}) for any $N$ when setting $\xi =\xi _{N}$, as
required by the property ii) in section \ref{Section-TTbar}.

\resection{Appendix} The expressions given in this paper for the form
factors of the components of the energy-momentum tensor and for $T\bar{T}$
apply to the sinh-Gordon model seen as a perturbation of the Gaussian (C=1)
fixed point when the form factors on states containing an odd number of
particles are set to zero. In this case the limit $b=0$ corresponds to a
free massive boson with background charge $Q=0$. The form factors of $\Theta$
and $T\bar{T}$ behave as 
\begin{equation}
F_{2m}^\Theta\sim b^{2(m-1)}\,,\hspace{1cm}F_{2m}^{T\bar{T}}\sim b^{2(m-2)}
\label{freelimit}
\end{equation}
as $b\to 0$. As expected, the only finite non-vanishing form factors at $b=0$
are obtained for $2$ and $4$ particles, respectively; there are however
infinities at $0$ particles for $\Theta$ and at $0$ and $2$ particles for $T%
\bar{T}$ that need to be subtracted.

The usual finite trace operator with zero expectation value at $b=0$ is
defined through the subtraction\footnote{%
The limit $b\to 0$ is understood in the equations below.} 
\begin{equation}
\Theta_R=\Theta-\langle\Theta\rangle\,I\,\,.
\end{equation}
The divergences of ${T\bar{T}}$ are eliminated in the subtracted expression 
\begin{eqnarray}
T\bar{T}_R &=&T\bar{T}-a\,m^{-2}\,\partial^2\bar{\partial}%
^2\Theta-d\,m^2\,\Theta-e\,m^4\,I+\mathcal{F}  \notag \\
&=& \mathcal{K}+c\,\partial\bar{\partial}\Theta+\mathcal{F}\,,
\end{eqnarray}
where $a$, $d$ and $e$ are the (infinite at $b=0$) coefficients given in
section~4, and $\mathcal{F}$ is a finite part needed to ensure that the
asymptotic factorization (\ref{clusterttbar}) continue to hold for the
regularized operators after the subtraction of the term proportional to $%
\partial^2\bar{\partial}^2\Theta$. This implies that $F_{n}^{\mathcal{F}}$
vanishes for $n\neq 4$ and is determined by 
\begin{equation}
{Q}_{4}^{\mathcal{F}}=\langle \Theta \rangle ^{2}\, \frac{\sigma
_{1}^{2}\sigma _{3}^{2}}{\sigma_{4}^{2}}\, (\sigma _{1}\sigma _{3}\sigma
_{2}-\sigma _{3}^{2}-\sigma _{1}^{2}\sigma _{4})\,,
\end{equation}
so that 
\begin{equation}
F_{4}^{T\bar{T}_R}=\left( \frac{\pi m^{2}}{2}\right) ^{2}\frac{1}{\sigma _{4}%
}(\sigma _{2}^{2}+14\sigma _{1}\sigma _{3}-4\sigma _{4})\,\,.
\label{TTbarb0}
\end{equation}
If $F_{2}^{T\bar{T}_R}$ is set to zero choosing $c=0$, this is the only
non-vanishing form factor of $T\bar{T}_R$.

This result is easily compared with that for the matrix elements of $T(x)%
\bar{T}(x)-\Theta (x)\Theta (x)$ computed in free field theory with normal
ordering regularization. The only difference arises in the coefficient of
the term proportional to $\sigma _{1}\sigma _{3}$ in (\ref{TTbarb0}),
amounting to a contribution of the operator $\partial \bar{\partial}\phi ^{4}
$. This is precisely one of the derivative terms in (\ref{regularized}) left
unfixed by (\ref{clusterttbar}).


\newpage

\begin{thebibliography}{99}
\bibitem{BPZ}  A.A. Belavin, A.M. Polyakov and A.B. Zamolodchikov, Nucl.
Phys. B 241 (1984) 333.

\bibitem{Taniguchi}  A.B. Zamolodchikov, Advanced Studies in Pure
Mathematics 19 (1989) 641; Int. J. Mod. Phys. A 3 (1988) 743.

\bibitem{KW}  M. Karowski, P. Weisz, Nucl. Phys. B 139 (1978) 455.

\bibitem{Smirnov}  F.A. Smirnov, Form Factors in Completely Integrable
Models of Quantum Field Theory, World Scientific, 1992.

\bibitem{counting}  J.L. Cardy and G. Mussardo, Nucl. Phys. B 340 (1990) 387.

A. Koubek, Nucl. Phys. B 435 (1995) 703.

F. Smirnov, Nucl. Phys. B 453 (1995) 807.

M. Jimbo, T. Miwa, Y. Takeyama, Counting minimal form factors of the
restricted sine-Gordon model, math-ph/0303059.

\bibitem{ttbar}  G. Delfino and G. Niccoli, Nucl. Phys. B 707 (2005) 381.

\bibitem{seven}  G. Delfino and G. Niccoli, J. Stat. Mech. (2005) P04004.

\bibitem{Sasha}  A.B. Zamolodchikov, Expectation value of composite field $T%
\bar{T}$ in two-dimensional quantum field theory, hep-th/0401146.

\bibitem{BM}  V.A. Belavin and O.V. Miroshnichenko, JETP Letters 82 (2005)
775.

\bibitem{LeClair}  A. LeClair, Phys. Lett. B 230 (1989) 103.

\bibitem{Smirnovreduction}  F.A. Smirnov, Nucl. Phys. B 337 (1990) 156.

\bibitem{RS}  N.Yu. Reshetikhin and F.A. Svirnov, Comm. Math. Phys. 131
(1990) 157.

\bibitem{BLc}  D. Bernard and A. LeClair, Nucl. Phys. B 340 (1990) 721.

\bibitem{DF}  Vl.S. Dotsenko and V.A. Fateev, Nucl. Phys. B240 (1984) 312.

\bibitem{ZZliouville}  A.B. Zamolodchikov and Al.B. Zamolodchikov, Nucl.
Phys. B 477 (1996) 577.

\bibitem{ZZ}  A.B. Zamolodchikov and Al.B. Zamolodchikov, Ann. Phys. 120
(1979) 253.

\bibitem{cth}  A.B. Zamolodchikov, Sov. J. Nucl. Phys. 46 (1987) 1090.

\bibitem{CT}  S. Coleman and H.J. Thun, Comm. Math. Phys. 61 (1978) 31.

\bibitem{Goebel}  C.J. Goebel, Prog. Theor. Phys. Suppl. 86 (1986) 261.

\bibitem{TBA}  Al.B. Zamolodchikov, Nucl. Phys. B 342 (1990) 695.

\bibitem{KlM}  T.R. Klassen and E. Melzer, Nucl. Phys. B 338 (1990) 485.

\bibitem{YL}  C.N. Yang and T.D. Lee, Phys. Rev. 87 (1952) 404; T.D. Lee and
C.N. Yang, Phys. Rev. 87 (1952) 410.

\bibitem{Fisher}  M.E. Fisher, Phys. Rev. Lett. 40 (1978) 1610.

\bibitem{Cardy}  J.L. Cardy, Phys. Rev. Lett. 54 (1985) 1354.

\bibitem{CMyanglee}  J.L. Cardy and G. Mussardo, Phys. Lett. B 225 (1989)
275.

\bibitem{AK}  I.Ya. Arefyeva and V.E. Korepin, Pis'ma Zh. Eksp. Teor. Fiz.
20 (1974) 680.

\bibitem{VG}  S.N. Vergeles and V.M. Gryanik, Yyad. Fiz. 23 (1976) 1324.

\bibitem{STW}  B. Schroer, T.T. Truong and P.H. Weisz, Phys. Lett. B 63
(1976) 422.

\bibitem{FMS}  A. Fring, G. Mussardo and P. Simonetti, Nucl. Phys. B 393
(1993) 413;

\bibitem{KM}  A. Koubek and G. Mussardo, Phys. Lett. B 311 (1993) 193.

\bibitem{DSC}  G. Delfino, P. Simonetti and J.L. Cardy, Phys. Lett. B 387
(1996) 327.

\bibitem{Luky}  S. Lukyanov, Mod. Phys. Lett. A 12 (1997) 2543.

\bibitem{MS}  G. Mussardo and P. Simonetti, Int. J. Mod. Phys. A9 (1994)
3307.

\bibitem{Cardycth}  J.L. Cardy, Phys. Rev. Lett. a60 (1998) 2709.

\bibitem{AlyoshaSG}  Al.B. Zamolodchikov, Int. J. Mod. Phys. A 10 (1995)
1125.

\bibitem{Koubek}  A. Koubek, Nucl. Phys. B 428 (1994) 655. 
%\bibitem{GiulianoTesi} G. Niccoli, Descendant operators in massive integrable quantum field 
%theories, SISSA PhD thesis, Trieste, November 2005.

\bibitem{AlyoshaYL}  Al. B. Zamolodchikov, Nucl. Phys. B 348 (1991) 619.

\bibitem{FFLZZ}  V. Fateev, D. Fradkin, S. Lukyanov, A. Zamolodchikov, Al.
Zamolodchikov, Nucl. Phys. B 540 (1999) 587.

\bibitem{BS}  P. Baseilhac and M. Stanishkov, Phys. Lett. B 554 (2003) 217;
Nucl. Phys. B 612 (2001) 373.

\bibitem{tim}  Al.B. Zamolodchikov, Nucl. Phys. B 358 (1991) 524.
\end{thebibliography}
\end{document}